\documentclass[letterpaper]{mn2e}
\usepackage{graphicx}
\usepackage{psfig}
\usepackage{times}
\usepackage{amsmath}
\usepackage{amssymb}

\title[Oscillations of vertically integrated relativistic tori I.]
{Oscillations of vertically integrated relativistic tori --
I. Axisymmetric modes in a Schwarzschild spacetime}

\author[Rezzolla, Yoshida and Zanotti]
	{Luciano Rezzolla$^{(1),(2)}$,
	Shin'ichirou Yoshida$^{(1)}$ and Olindo Zanotti$^{(1)}$ \\
								\\
	$^1$SISSA, International School for Advanced Studies,	
        Via Beirut, 2 34014 Trieste, Italy			\\
	$^2$INFN, Sezione di Trieste, Via Valerio, 2 34127 	
	Trieste, Italy
	}

\begin{document}

\maketitle
\pagerange{\pageref{firstpage}--\pageref{lastpage}}
\pubyear{2003}

\label{firstpage}

\begin{abstract}
	This is the first of a series of papers investigating the
	oscillation properties of relativistic, non-selfgravitating tori
	orbiting around a black hole. In this initial paper we consider
	the axisymmetric oscillation modes of a torus constructed in a
	Schwarzschild spacetime. To simplify the treatment and make it as
	analytical as possible, we build our tori with vertically
	integrated and vertically averaged quantities, thus transforming
	the eigenvalue problem into a set of coupled ordinary
	differential equations. The tori are also modeled with a number
	of different non-Keplerian distributions of specific angular
	momentum and we discuss how the oscillation properties change
	when different distributions of angular momentum are
	considered. Our investigation progresses by steps. We first
	consider a local analysis in Newtonian gravity and determine the
	properties of acoustic wave propagation within these objects, as
	well as the relations between acoustic and epicyclic
	oscillations. Next, we extend the local analysis to a general
	relativistic framework. Finally, we perform a global analysis and
	determine both the eigenfunctions and the eigenfrequencies of the
	axisymmetric oscillations corresponding to the $p$ modes of
	relativistic tori. These behave as sound waves globally trapped
	in the torus and possess eigenfrequencies appearing in the simple
	sequence 2:3:4:$\ldots$, independently of the distribution of
	angular momentum considered. The properties of the modes
	investigated here are in good agreement with those observed in
	recent numerical simulations and could have a number of different
	applications. In X-ray binary systems containing a black hole
	candidate, for instance, $p$-mode oscillations could be used to
	explain the harmonic relations in the high frequency
	quasi-periodic oscillations observed. In systems comprising a
	massive torus orbiting a black hole, on the other hand, $p$-mode
	oscillations could be used to explain the development or the
	suppression of the runaway instability.
\end{abstract}


\date{Accepted 0000 00 00.
      Received 0000 00 00.}


\section{Introduction}
\label{intro}

	Waves and normal-mode oscillations in stars have long since
attracted attention and a vast literature, investigating them both in
Newtonian (see, for instance, Cox 1980, or Unno et al., 1989) and in
general relativistic regimes (see Stergioulas 1998, for a review), is now
available. On the other hand, waves and normal-mode oscillations in
geometrically thin discs around compact objects have been studied much
less, both within Newtonian gravity (see Kato 2001 for a review) and
within a relativistic framework (Okazaki et al., 1987; Perez et al.,
1997; Silbergleit et al., 2001; Kato 2001, Rodriguez et al. 2002). The
literature is even more scarce when one considers geometrically thick
discs, which so far have have been investigated mostly in connection with
their stability properties both in Newtonian gravity (Papaloizou and
Pringle 1984, 1985; Blaes 1985) and in general relativity (Kojima,
1986). An explanation for why discoseismology has not yet reached the
development and the level of sophistication which is now possible in
asteroseismology is due, at least in part, to the fact that only recently
accretion discs have been recognized as fundamental astrophysical
objects, present in various forms at all scales. Nowdays, however,
periodic and quasi-periodic variations are currently observed in
different classes of astrophysical objects containing accretion discs.
Although many different models have been proposed for the interpretation
of the rich phenomenology associated with these quasi-periodic
oscillations (QPOs), there seems not to be yet a widely accepted
mechanism for most of the observed sources (see van der Klis, 2000 for a
review). Clearly, a systematic investigation of the oscillation
properties of discs could shed some light on this.

	As in stars, oscillation modes in discs are, in general, the
consequence of restoring forces responding to perturbations and these
offer a way for classifying oscillations. In accretion discs, in
particular, a first restoring force is given by the centrifugal force,
which is responsible for the appearance of the so called {\it inertial
waves}, tightly related to the orbital motion of the disc and hence to
epicyclic oscillations.  A second restoring force is offered by pressure
gradients and the oscillations that are produced in this way are related
to $p$ modes and have close connections with the propagation of {\it
sound waves} in the perturbed fluid. A third restoring force is the
gravitational field in the direction orthogonal to the orbital plane. If
a portion of the disc is perturbed in the vertical direction, in fact,
the vertical component of the gravitational field will produce a harmonic
oscillation across the equatorial plane with oscillation frequency equal
to the orbital frequency. These oscillations are related to {\it
corrugation waves} (see Kato, Fukue and Mineshige, 1998, for an overview
on disc oscillations).

	Two complementary approaches have been followed traditionally for
studying the perturbations of equilibrium models of geometrically thin
accretion discs.  The first one is a {\it local} approach, and it has
been used extensively to investigate the propagation of waves with the
inclusion of many contributing physical effects, such as buoyancy,
stratified atmospheres and magnetic fields (see, among the others, Lubow
and Ogilvie, 1998). Being local, these approaches derive dispersion
relations in which the frequency of the perturbation is expressed as a
function of the spatial position within the object and as the linear
combination of different contributions, each related to the different
physical effect (Kato, 2001). The second approach is a {\it global} one
and it is based on the fact that, under suitable conditions, Eulerian
perturbations to all physical quantities can be expressed as an
eigenvalue problem, the eigenvalue being the frequency of oscillation. In
practice, and for the simplest case, this amounts in solving a second
order partial differential equation once appropriate boundary conditions
are provided (see Ipser and Lindblom, 1992; Silbergleit et al., 2001;
Nowak and Wagoner, 1991).
 
	This paper is devoted to both a local and a global perturbative
analysis of axysimmetric modes of oscillation of relativistic tori in the
Cowling approximation (i.e. in an approximation in which the
perturbations of the spacetime are neglected; Cowling 1941). Fluid tori
differ from geometrically thin discs mostly in having equilibrium
configurations in which the orbital motion is intrinsically non-Keplerian
and in which pressure gradients play an important role, giving rise to an
extended vertical structure. As a result, a consistent investigation of
these configurations which would account for the coupling of the
oscillations in the radial and vertical directions requires necessarily a
two-dimensional treatment involving a set of partial differential
equations. While this problem is solvable with presently available
techniques (this has indeed already been done for relativistic rotating
stars; Yoshida and Eriguchi, 1997), it would require a massive use of
expensive numerical calculations leaving little room to a physical
interpretation. Furthermore, in contrast with the case of relativistic
stars, the eigenvalue problem for relativistic tori is in great part
unsolved and resorting to a fully numerical solution at this stage would
prevent one from appreciating most of the basic physics behind the
oscillation modes of these objects. For this reason, we will here
concentrate on a simpler model for the torus which can be handled in
great part analytically, and postpone the use of the two-dimensional
fully numerical analysis to a subsequent work.

	The main simplification in the models discussed here is that the
vertical structure of the tori is accounted for by integrating the
relevant quantities along the direction perpendicular to the equatorial
plane. Doing so removes one spatial dimension from the problem, which can
then be solved integrating simple ordinary differential equations. The
local approach, in particular, will allow us to derive the local
dispersion relation obeyed by the oscillations in relativistic
non-Keplerian discs, while the global approach will provide us with the
eigenfunctions and eigenfrequencies of the system.

	There are several motivations behind this study. Firstly, we want
to extend the relativistic discoseismology analysis carried so far for
thin discs to systems having a non negligible contribution coming from
pressure gradients. Secondly, we intend to interpret and clarify some of
the numerical results found in the time evolution of ``toroidal neutron
stars'', i.e. compact and massive tori orbiting a Schwarzschild black
hole (Zanotti, Rezzolla \& Font, 2003). Thirdly, we want to assess the
possible connections between the oscillation modes of relativistic tori
and the rich X-ray phenomenology observed in QPOs. Finally, we want to
investigate the possibility that the axisymmetric oscillations of thick
discs could provide a criterion for the development or the suppression of
the runaway instability (Abramowicz et al., 1983).

	The plan of the paper is as follows: in Section~\ref{NT} we
briefly review the local analysis of oscillation modes in vertically
integrated Newtonian tori. Section~\ref{GRT}, on the other hand,
introduces the basic assumptions and equations employed in the definition
of our general relativistic, vertically integrated torus. These equations
will then be used to study axisymmetric oscillations both locally, in
Section~\ref{GRT_la}, and globally, in Section~\ref{GRT_ga}. We will
first consider configurations with constant distributions of specific
angular momentum, and subsequently distributions of specific angular
momentum that are either linear or power-laws in the cylindrical radial
coordinate. Finally, Section~\ref{conclusions} contains our conclusions
and the prospects for further investigations. Hereafter Greek indices are
taken to run from 0 to 3 and Latin indices from 1 to 3; unless stated
differently, we will use units in which $G = c = M_{\odot} = 1$.

\section{Newtonian Tori: a Local Analysis}
\label{NT}

	We here briefly present a local analysis of the oscillation modes
of vertically integrated tori within a Newtonian framework. Some of the
results presented in this Section have already been discussed in the
literature, but will serve as a useful reference for the general
relativistic treatment presented in Section~\ref{GRT_la}, and will help
in the physical interpretation of the relativistic results.

\subsection{Assumptions and Equations}

	Consider an extended perfect fluid configuration orbiting a
central object which is the only source of the gravitational potential
(i.e. the orbiting fluid is non-selfgravitating). Introduce now a
cylindrical coordinate system, ($\varpi, \phi$, $z$) whose origin is at
the center of the central object and whose $z$-axis is oriented along the
direction of the orbital angular momentum vector. 

	A simplified description of this system can be obtained by
removing the dependence of the various physical quantities on the
vertical coordinate $z$. Mathematically, this is done by integrating the
relevant physical quantities along the vertical direction. Physically,
this corresponds to collapsing the vertical structure of the torus onto
the equatorial plane, but is quite {\it different} from just considering
an equatorial slice of the vertically extended torus.

	Using this approach, it is possible to define the {\it vertically
integrated} pressure $P$
\begin{equation}
\label{vert_p}
	P(\varpi) \equiv \int_{-H}^{H}p dz\ ,
\end{equation}
and the vertically integrated rest-mass density $\Sigma$
\begin{equation}
\label{vert_Sigma}
	\Sigma (\varpi) \equiv \int_{-H}^{H}\rho dz\ ,
\end{equation}
where $H=H(\varpi)$ is the local ``thickness'' of the torus. An equation
of state (EOS) for the fluid in the torus needs to be specified and we
find it convenient to use here a simple barotropic EOS of polytropic
type, i.e.
\begin{equation}
\label{poly_0}
p = k \rho^{\gamma}\ , 
\end{equation}
where $k$ and $\gamma\equiv d\ln p/d\ln\rho$ are the polytropic constant
and the adiabatic index, respectively. The vertically integrated
quantities in (\ref{vert_p}) and (\ref{vert_Sigma}) need also to be
related through an ``effective'' equation of state and this can be done
if we define an ``effective'' adiabatic index $\Gamma\equiv d\ln
P/d\ln\Sigma$, so that
\begin{equation}
\label{poly_1}
P = {\cal K} \Sigma^{\Gamma}\ , 
\end{equation}
with the constants ${\cal K}$ and $\Gamma$ playing the role of the
polytropic constant and of the adiabatic index, respectively. Note that
while equation (\ref{poly_1}) mimics a polytropic equation of state, it
does not represent a vertically integrated polytropic equation of state
(unless, of course, the adiabatic index is equal to one). More
importantly, the adiabatic index $\Gamma$ is {\it not} constant but
depends both on $\varpi$ and $z$. This complication, however, can be
removed if one assumes that both the pressure and the rest-mass density
have a weak dependence on height, so that they can be accurately
expressed in terms of their values at the equator, i.e.
\begin{eqnarray}
\label{poly_2}
&&p = p(\varpi,z) \approx p_0 \equiv p(\varpi,z=0) \ , 
\\ \nonumber \\
&&\rho = \rho(\varpi,z) \approx \rho_0 \equiv \rho(\varpi,z=0) \ . 
\end{eqnarray}
Using this assumption, $P \approx 2 H p_0$ and $\Sigma \approx 2 H
\rho_0$, so that \hbox{$\Gamma \approx d\ln p_0/d\ln\rho_0=\gamma$} and
equation (\ref{poly_1}) can effectively be written as
\begin{equation}
\label{poly_3}
P = k \Sigma^{\gamma}\ . 
\end{equation}
In all of the calculations reported here we have used $\Gamma=4/3$, but
the results do not change qualitatively when different polytropic indices
are used.

	Being in circular non-Keplerian motion with angular velocity
$\Omega$, the torus will have velocity components only in the $\varpi$
and in the $\phi$-directions, with {\it vertical averages} given by
\begin{equation}
\label{vert_u}
	U (\varpi) \equiv \frac{1}{2H}\int_{-H}^{H}v^{\varpi} dz\ ,
\end{equation}
and
\begin{equation}
\label{vert_w}
	W (\varpi) \equiv \frac{1}{2H}\int_{-H}^{H}v^{\phi} dz\ .
\end{equation}

	With these assumptions, the dynamics of the torus is fully
described by a set of equations comprising the Euler equations in the
$\varpi$ and $\phi$-directions (Shu, 1992)
\begin{eqnarray}
\label{equilibrium_2}
&& \partial_t U + U\partial_\varpi U + \frac{W}{\varpi}\partial_\phi U
	-\frac{W^2}{\varpi} = -\frac{1}{\Sigma}\partial_\varpi P -
	\partial_\varpi\Psi\ , 
\\ \nonumber \\
\label{equilibrium_3}
& & \partial_tW + U\partial_\varpi W + \frac{W}{\varpi}\partial_\phi W
	+ \frac{UW}{\varpi} = 
\nonumber \\
&& \hskip 4.0truecm	-\frac{1}{\Sigma}\frac{1}{\varpi}
	\partial_\phi P - \frac{1}{\varpi}\partial_\phi\Psi \ , 
\end{eqnarray}
the equation for the conservation of mass 
\begin{equation}
\label{equilibrium_1}
\partial_t\Sigma + \partial_\varpi(\Sigma U) + \frac{1}{\varpi}\Sigma
	U + \frac{1}{\varpi}\partial_\phi(\Sigma W) = 0 \ , 
\end{equation}
and the Poisson equation for a vertically integrated gravitational
potential $\Psi$
\begin{equation}
\label{equilibrium_4}
\nabla^2 \Psi = 4\pi {\cal S} \ ,
\end{equation}
where ${\cal S}$ is the vertically integrated rest-mass density of the
central object, only source of the gravitational potential.

	 Stationary and axisymmetric models for the tori can be built
after setting to zero the terms in
(\ref{equilibrium_2})--(\ref{equilibrium_1}) involving derivatives with
respect to the $t$ and $\phi$-coordinates. The solutions obtained in this
way represent the background equilibrium solutions over which harmonic
Eulerian perturbations of the type
\begin{equation}
\label{harmonic}
\left(
	\begin{array}{c}
\delta U\\ \nonumber \\
\delta W\\ \nonumber \\
\delta q
	\end{array}
\right) 
	\sim e^{-i\sigma t + ik_{\varpi}\varpi} \ ,
\eqno{(14)}
\end{equation}
can be introduced. Note that Eulerian perturbations are indicated with a
``$\delta$'' to distinguish them from the corresponding Lagrangian
perturbations, that we will instead indicate with a ``$\Delta$''. In
expression (\ref{harmonic}), $\sigma$ is the frequency of the
perturbation (in general a complex number), $k_{\varpi}$ is the
wavenumber in the $\varpi$-direction (hereafter simply $k$), and $\delta
q\equiv \delta P/\Sigma$ has been introduced to describe the
perturbation in the pressure.

	We have here neglected the perturbations in the gravitational
potential and therefore set $\delta \Psi = 0$. This is referred to as the
Cowling approximation (Cowling 1941) and for a fluid configuration which
is non-selfgravitating (i.e. one whose gravitational potential is such
that $\Psi_{\rm fluid} + \delta \Psi_{\rm fluid}=0$), the Cowling
approximation is actually an exact description of the pulsations (Ipser
and Lindblom, 1992).

	Note that the harmonic spatial dependence expressed in
(\ref{harmonic}) is the signature of the local approach and is valid as
long as the wavelength of the perturbations considered is smaller than
the lengthscale of the radial variations in the equilibrium configuration
(this is basically the condition for the WKB approximation) or, stated
differently, that
\begin{equation}
\setcounter{equation}{15}
\label{cf}
k \gg \frac{1}{P}\frac{d P}{d \varpi} 
	 \sim \frac{1}{P}\frac{d P}{d \varpi}
	\sim \frac{1}{L}\ .
\end{equation}

	Introducing the perturbations (\ref{harmonic}) in the equilibrium
model given by equations (\ref{equilibrium_2})--(\ref{equilibrium_4}) and
retaining only the first-order terms, we derive the following
perturbation equations (Shu, 1992)
\begin{eqnarray}
\label{linearized_1}
&& i\sigma\delta U + 2\Omega\delta W = i k \delta q \ ,\\ \nonumber \\
\label{linearized_2}
&& (2\Omega +
	\varpi\partial_\varpi\Omega)i\delta U +
	\sigma\delta W
	= 0 \ ,\\ \nonumber \\
\label{linearized_3}
&& i\sigma \delta q - ik c^2_s \delta U = 0 \ ,
\end{eqnarray}
where $c^2_s \equiv dP/d\Sigma=\Gamma P/\Sigma$ is the local sound
speed. We can now cast equations
(\ref{linearized_1})--(\ref{linearized_3}) into a simple matrix form as
\begin{equation}
\label{real_matrix}
\left(
\begin{array}{ccc}
\sigma & 2\Omega & -k\\	
	\\
\displaystyle\frac{\kappa_{\rm r}^2}{2\Omega} & \sigma & 0\\
	\\
	k c^2_s & 0 & -\sigma
\end{array}
\right) 
\left(
	\begin{array}{c}
	i\delta U\\ 
	\\
	\delta W\\ 
	\\
	i \delta q		
	\end{array}
\right)
		=  0 \ ,
\end{equation}
where $\kappa_{\rm r}$ is the {\it epicyclic frequency} in the radial
direction\footnote{In principle, an epicyclic oscillation takes places
also for motions away from the background orbital plane. The
corresponding frequencies are the vertical epicyclic frequencies but
these will not be considered here.} and is defined as
\begin{equation}	
\label{newton-epicyclic}
\kappa_{\rm r}^2 \equiv 2\Omega\left(2\Omega +
	\varpi\frac{d\Omega}{d\varpi}\right)\ . 
\end{equation}	
Note that the radial epicyclic frequency is equal to the orbital
frequency for Keplerian orbital motion.  i.e. $\kappa_{\rm r}^2 =
\Omega^2$ for $\Omega \sim \varpi^{-3/2}$, and is zero for motions with
constant specific angular momentum, i.e. for $\ell\equiv \Omega
\varpi^2={\rm const}$.

	Being a homogeneous linear system, a non-trivial solution to the
set of equations (\ref{real_matrix}) exists if the determinant of the
coefficients matrix is equal to zero, which then provides the dispersion
relation
\begin{equation}
\label{dispersion-newton}
\sigma^2 = \kappa_{\rm r}^2 + k^2 c^2_s \ .
\end{equation}

	This approximate form of the dispersion relation was first
applied to waves in accretion discs by Okazaki et al. (1987) and then
reconsidered by several authors in more general situations (see Nowak and
Wagoner, 1992; Ipser, 1994; Silbergleit et al., 2001).  The two terms in
the dispersion relation (\ref{dispersion-newton}) are most easily
interpreted when considered separately. To this scope, consider a torus
composed of collisionless particles and with specific angular momentum
increasing outwards. A fluid element which is infinitesimally displaced
from its equilibrium orbit but conserves its angular momentum unchanged,
will start oscillating in the radial direction due to a restoring
centrifugal force. These oscillations are called {\it inertial
oscillations} and their frequency is the radial epicyclic frequency
$\kappa_{\rm r}(\varpi)$. In compressible fluids, on the other hand, a
restoring force due to pressure gradients is also present and is
responsible for {\it acoustic oscillations} with frequency $k c_s$. Both
of these terms contribute to the right-hand-side of the dispersion
relation (\ref{dispersion-newton}) and are collectively referred to as
{\it ``inertial-acoustic waves''}. Following a standard convention (Kato,
2001), we will here identify the high frequency modes $(\sigma^2\gtrsim
\kappa_{\rm r}^2)$ with the $p$ modes, or inertial-acoustic modes (Kato
and Fukue, 1980) of the vertically integrated torus.

\section{Relativistic Tori: Assumptions and Equations}
\label{GRT}

	As for the Newtonian case, we will here assume that the torus
does not contribute to the spacetime metric, which we will take to be
that external to a Schwarzschild black hole. This is the simplest, yet
non-trivial metric to consider and, more importantly, it allows for a
direct comparison with the numerical calculations of Zanotti et al.
(2003) (The extension of this work to the Kerr metric will be presented
in a future work; Rezzolla and Yoshida, 2003). Furthermore, since we are
really interested in the portion of the spacetime in the vicinity of the
equatorial plane (i.e. for values of the spherical angular coordinate
$|\theta - \pi/2| \ll 1$), we will write the Schwarzschild metric in
cylindrical coordinates and retain the zeroth-order terms in the ratio
$(z/r)$ (Novikov and Thorne, 1973). In this case, the line element takes
the form
\begin{equation}
\label{eqtrl_metric}
ds^2 = -e^{2\nu(\varpi)}dt^2 + e^{2\lambda(\varpi)} d\varpi^2 
	+ dz^2 + \varpi^2 d\phi^2\ ,
\end{equation}
where 
\begin{equation}
e^{2\nu(\varpi)} = 1 - \frac{2M}{\varpi}=e^{-2\lambda(\varpi)} \ . 
\end{equation}

	The basic equations to be solved in order to construct
equilibrium models are the continuity equation for the rest-mass density
$\rho$
\begin{equation}
\label{cont_e}
\nabla_\alpha(\rho u^\alpha)=0 \ ,
\end{equation}
and the conservation of energy-momentum
\begin{equation}
\label{set}
\nabla_\alpha T^{\alpha\beta}=0 \ , 
\end{equation}
where the symbol $\nabla$ refers to a covariant derivative with respect
to the metric (\ref{eqtrl_metric}). In equation (\ref{set}),
$T^{\alpha\beta} \equiv (e+p) u^{\alpha}u^{\beta} + p g^{\alpha \beta}$
are the components of the stress-energy tensor of a fluid with isotropic
pressure $p$ and total energy density $e$. Since we are dealing with a
curved background spacetime, but we want a close comparison with
Newtonian expressions, it is useful to introduce an orthonormal tetrad
carried by the local static observers and defined by the one-form basis
\begin{equation}	
{\boldsymbol \omega}^{\hat{t}} 	
	= e^\nu {\boldsymbol d}t\ ,\quad
{\boldsymbol \omega}^{\hat{\varpi}} 
	= e^\lambda {\boldsymbol d}\varpi\ ,\quad
{\boldsymbol \omega}^{\hat{z}} 
	= {\boldsymbol d}z\ ,\quad
{\boldsymbol \omega}^{\hat{\phi}} 
	= \varpi {\boldsymbol d}\phi\ .
\label{1-form}
\end{equation}
In this frame, the components of the fluid four-velocity are denoted by
$u^{\hat{\mu}}$ (with $\mu=t, \varpi,z,\phi$), and the 3-velocity
components are then given by
\begin{equation}
v^{\hat{i}}   = \frac{u^{\hat{i}}}
{u^{\hat{t}}} = \frac{\omega^{\hat{i}}_{\alpha} u^{\alpha}}
	{\omega^{\hat{t}}_{\alpha}u^{\alpha}} \ , 
\hspace{1cm} i=\varpi, z, \phi \ .
\end{equation}

	In analogy with the Newtonian case presented in Section~\ref{NT},
we use the vertically integrated quantities defined in
(\ref{vert_p})--(\ref{vert_Sigma}) and assume that they obey an effective
polytropic equation of state of the type (\ref{poly_1}). Enforcing the
conditions of hydrostatic equilibrium and of axisymmetry simplifies the
hydrodynamical equations considerably, reducing them to Bernoulli-type
equations (Kozlowski et al., 1978)
\begin{equation}
\label{bernoulli}
\frac{\nabla_i p}{e+p} = - \nabla_i \ln (u_t) +
	\frac{\Omega \nabla_i \ell}{1- \Omega \ell} \ ,
\end{equation} 
where $\ell \equiv -u_{\phi}/u_t$ is the specific angular momentum
(i.e. the angular momentum per unit energy). Simple algebraic
manipulations show that the only relevant component of equations
(\ref{bernoulli}), i.e. the radial one, can be written explicitly as
\begin{equation}
\label{quasi}
\frac{\partial_\varpi p}{e+p}=-
	\frac{e^{2\nu} \partial_{\varpi}\nu - 
	\Omega^2\varpi}{e^{2\nu} - \Omega^2\varpi^2} 
	\ .
\end{equation}
which represents a generic condition for the existence of hydrostatic
equilibrium of a fluid configuration orbiting around a Schwarzschild
black hole. Note also that a number of cancellations have removed any
dependence on the spatial derivative of the angular velocity from the
right-hand-side of equation (\ref{quasi}).

	Being interested in vertically integrated expressions, we need to
integrate both sides of equation (\ref{quasi}). Interestingly, in the
spacetime (\ref{eqtrl_metric}) the right-hand-side of (\ref{quasi}) is a
function of the cylindrical radial coordinate only, and hence its
vertical integration simply yields a new multiplicative factor $2H$. The
left-hand-side, on the other hand, cannot be managed unless a simplifying
assumption is made and this amounts to consider that the total enthalpy
$(e+p)$ is only a weak function of the $z$-coordinate so that,
effectively, it is possible to substitute it with its vertically averaged
value $({\bar e} + {\bar p})$, where
\begin{equation}
e(\varpi,z) + p(\varpi,z) \approx {\bar e}(\varpi) + {\bar p}(\varpi)
	\equiv \frac{1}{2H} \int^H_{-H} (e+p)dz \ .
\end{equation}	
With this assumption, the vertically integrated equation (\ref{quasi})
can be written as
\begin{equation}
\label{vert_eq}
\frac{1}{E+P}\frac{d P}{d\varpi} 
	= - \frac{\varpi}{\varpi - 2M - \Omega^2 \varpi^3}
	\left(
	\frac{M}{\varpi^2} - \Omega^2 \varpi 
	\right)\ ,
\end{equation}	
where $E$ is the vertically integrated energy density
\begin{equation}
\label{vert_e}
	E \equiv \int_{-H}^{H}e dz = \frac{P}{\Gamma - 1} + \Sigma\ 
	\approx \frac{P}{\gamma - 1} + \Sigma\ .
\end{equation}
	We next introduce perturbations in the velocity and pressure with
a harmonic time dependence of the type
\begin{equation}
\label{harmonic_r}
\left(\begin{array}{c}
\delta V^{\hat{\varpi}}\\ \nonumber \\
\delta V^{\hat{\phi}}\\ \nonumber \\
\delta Q
	\end{array}
\right) \sim e^{-i\sigma t} \ , 
\eqno(33)
\end{equation}
where $\delta Q\equiv\delta P/(E+P)$, and where we have defined the
averaged velocity perturbations as
\begin{equation}
\setcounter{equation}{34}
\label{vert_g1}
\delta V^{\hat{\varpi}} \equiv
	\frac{1}{2H}\int_{-H}^{H}\delta v^{\hat{\varpi}}dz \ , \qquad
\delta V^{\hat{\phi}} \equiv
	\frac{1}{2H}\int_{-H}^{H}\delta v^{\hat{\phi}}dz \ . 
\end{equation}

	After perturbing the hydrodynamical equations
(\ref{cont_e})--(\ref{set}) and taking the vertically integration of the
resulting equations, we derive the following set of equations
\begin{eqnarray}
\label{euler-r}
&&      i\sigma e^{\nu-\lambda}\delta V^{\hat{\varpi}}
	+ 2e^{\nu-2\lambda}\Omega 
	\left(1+
	\frac{\varpi}{E+P}\frac{d P}{d \varpi}\right) 
	\delta V^{\hat{\phi}} +
\nonumber\\
&& \hskip 4.0truecm	
	Ae^{-2\lambda} \frac{d (\delta Q)}{d \varpi}  = 0 \ , 
\\\nonumber\\
\label{euler-phi}
&&      i\sigma \delta V^{\hat{\phi}}
	- \left(\frac{d \Omega}{d \varpi}+\frac{2}{\varpi}\Omega- 
	2\Omega\frac{d \nu}{d \varpi}\right)
	\varpi e^{-\lambda} \delta V^{\hat{\varpi}} +
\nonumber\\
&& \hskip 4.0truecm
	i\sigma A\Omega e^{-2\nu} \varpi \delta Q = 0 \ ,
\\\nonumber\\
\label{cont_gr}
&&      i\sigma \left(\frac{E +P}{\Gamma P}\right) \delta Q 
	- e^{\nu-\lambda} \frac{d (\delta V^{\hat{\varpi}})}{d \varpi}
	-i\sigma\frac{\varpi e^{\nu}}{A}\Omega \delta V^{\hat{\phi}} -
\nonumber\\
&& \hskip 0.5truecm
	\left[2\frac{d \nu}{d \varpi} + 
	\frac{1}{\varpi} - 
	\frac{1}{2A}\frac{d A}{d \varpi}
	+\frac{1}{\Gamma P}\frac{d P}{d \varpi}\right]
	e^{\nu-\lambda}\delta V^{\hat{\varpi}} = 0\ , 
\nonumber\\
\end{eqnarray}
where $A\equiv 1/(u_t)^2$. Equations (\ref{euler-r})--(\ref{cont_gr})
represent the $\varpi$ and $\phi$-components of the perturbed
relativistic Euler equations as well as the perturbed continuity
equation.

\section{Perturbations of Relativistic Tori: a Local Analysis}
\label{GRT_la}

	A local analysis of the perturbed hydrodynamical equations
(\ref{euler-r})--(\ref{cont_gr}) can be performed after introducing a
harmonic radial dependence of the type $\delta Q, \delta
V^{\hat{\varpi}}, \delta V^{\hat{\phi}} \sim \exp{(i k \varpi)}$, where
$k$ is the radial wavenumber and, again, we will assume that $\lambda =
2\pi/k \ll L$. Simple considerations allow now to remove some of the
terms in the continuity equation (\ref{cont_gr}), thus simplifying it
further. In particular, it is easy to realize that the third and fourth
terms in equation (\ref{cont_gr}) are very small as compared to the first
and second ones, to which they are similar in nature. More specifically,
both the second and the fourth terms involve the radial velocity
perturbation but with coefficients proportional to $k$ and $1/L$,
respectively. Assuming that the condition (\ref{cf}) is satisfied, it is
then reasonable to drop the fourth term of equation
(\ref{cont_gr}). Similarly, the ratio between the third and the first
term can be approximated as
\begin{equation} 
\label{ratio_3_1}
\frac{\delta V^{\hat{\phi}}}{\delta Q}
	\frac{\Gamma P \varpi \Omega e^{\nu}} {\left( E+P \right) A}
	\sim {\cal O}\left(\frac{\delta V^{\hat{\phi}}}{\delta Q}\right) 
	\times {\cal O}(|\mbox{typical	velocity}|^3)\ ,  
\end{equation} 
where we have approximated the first term as $\sim \delta P/(\Gamma
P)\sim \delta Q/c^2_s$ and the third one as $\sim (\varpi\Omega)\delta
V^{\hat{\phi}}$. While $(\delta V^{\hat{\phi}}/\delta Q)$ is the ratio of
two perturbed quantities and of order unity, the typical velocity of is
of the order of the orbital velocity at the marginally stable circular
orbit (i.e. $\sim 1/\sqrt{6}$) so that, overall, the ratio in
(\ref{ratio_3_1}) is much smaller than unity and the third term in
equation (\ref{cont_gr}) can therefore also be dropped.  Finally, it
should be noted that the third term on the left hand side of equation
(\ref{euler-phi}) can also be discarded, as it would introduce a purely
imaginary part in the dispersion relation (\ref{dispersion-gr}). This
term, which is due a time derivative of the pressure in the Euler
equations, has a purely relativistic origin and has been found to provide
negligible contributions in the numerical solution of the system
(\ref{euler-r})-(\ref{cont_gr}).
	
	We can now introduce quantities that are more directly related to
the ones used in the Newtonian example of Section~\ref{NT}, namely
\begin{equation}
\delta U \equiv i\delta V^{\hat{\varpi}} \ , 
	\qquad \delta W\equiv \delta V^{\hat{\phi}} \ , 
\end{equation}
so that the linearized perturbation equations in the unknowns $\delta U$,
$\delta W$ and $\delta Q$ can be written as a homogeneous linear system
\begin{equation}
{\cal M}(k,\sigma) 
\left(\begin{array}{c}
	\delta U \\ \nonumber \\ 
	\delta W \\ \nonumber \\ 
	\delta Q
\end{array}\right)
	= 0 \ ,
\eqno	(40)
\end{equation}
where the matrix of coefficients ${\cal M}$ depends both on $k$ and
$\sigma$. Imposing the determinant of ${\cal M}$ to be zero, we obtain
the dispersion relation
\begin{eqnarray}
\setcounter{equation}{41}
\label{dr_0}
&&      \sigma^3 - \sigma e^{-2\lambda}
	\biggl[2\Omega
	\left(2\Omega + \varpi\frac{d \Omega}{d \varpi}
	- 2\varpi\Omega\frac{d \nu}{d \varpi}\right)
	\biggl(1+
\nonumber \\
&& \hskip 0.5truecm 
	\left.
	\frac{\varpi}{E+P}\frac{d P}{d \varpi}\right) 
	+ e^{2\nu}\left(1-e^{-2\nu}\varpi^2\Omega^2\right) k^2
	\frac{\Gamma P}{E+P}\biggr] = 0 \ .
\nonumber \\
\end{eqnarray}
A non-trivial solution of (\ref{dr_0}) is given by
\begin{equation}
\label{dispersion-gr}
\sigma^2 = \kappa_{\rm r}^2 + 
	[e^{2(\nu-\lambda)} (1-e^{-2\nu}\varpi^2\Omega^2)]
	k^2 c^2_s \ ,
\end{equation}
where now $c^2_s \equiv dP/dE$ is the square of the relativistic sound
velocity in the vertically integrated disc. The dispersion relation
(\ref{dispersion-gr}) represents the relativistic generalization of
(\ref{dispersion-newton}) in which we have defined the relativistic
radial epicyclic frequency for an extended fluid object $\kappa_{\rm r}$
as
\begin{equation}
\label{gr_epicyclic}
\kappa_{\rm r}^2 \equiv 2 e^{-2\lambda} \Omega \left(
	2\Omega + \varpi\frac{d \Omega}{d \varpi} - 
	2\varpi\Omega\frac{d \nu}{d \varpi} \right)
	\left(1+\frac{\varpi}{E+P}\frac{d P}{d \varpi}
	\right)	\ .
\end{equation}

	A number of comments should now be made about this
expression. Firstly, while the definition (\ref{gr_epicyclic}) reduces to
the corresponding expression (\ref{newton-epicyclic}) when the Newtonian
limit is taken, it has a new important correction coming from pressure
gradients (cf. second term in the second round brackets, which is a
function of $\varpi$ only). Secondly, in those cases in which the
(radial) pressure gradients can be neglected (e.g. for very thin discs),
the fluid motion is essentially Keplerian and expression
(\ref{gr_epicyclic}) coincides with the relativistic radial epicyclic
frequency for a point-like particle. In this case, and only in this case,
(\ref{gr_epicyclic}) reduces to the expression for the radial epicyclic
frequency derived by Okazaki et al. (1987) that, for a Schwarzschild
spacetime in spherical coordinates, is
\begin{equation}
\label{epic_kep}
(\kappa_{\rm r})_{\rm Keplerian} = \sqrt{
	 \left(1-\frac{6M}{r}\right)\frac{G M}{r^3}} \ .
\end{equation}

	An important feature of the relativistic radial epicyclic
frequency (\ref{epic_kep}), and that distinguishes it from the
corresponding expression (\ref{gr_epicyclic}), is that such a function is
not monotone in $\varpi$ but has a maximum at a given radius, thus
indicating that in General Relativity oscillation modes can be trapped
near the inner edge of an accretion disc (Kato and Fukue, 1980). When
pressure gradients are taken into account, however, this ceases to be
true and the epicyclic frequency becomes monotonically decreasing with
radius (cf. Figure~\ref{epic_1})
\begin{table*}
\begin{minipage}{120mm}
\begin{center}
\begin{tabular}{ccccccccccccc}\hline  
\label{tab1}
Model 	& $\ell$-distribution
	& ${\tilde \ell}_{\rm c}$
	& $\alpha,\; \beta$
	& $q$
	& ${\widetilde \varpi}_{\rm in}$ 
	& ${\widetilde \varpi}_{\rm out}$ 
	& ${\widetilde \varpi}_{\rm cusp}$  
	& ${\widetilde \varpi}_{\rm max}$
	& $\Delta W_{\rm in}$  
	\\
	\hline \\
(c1)& const. 	& 3.90  & $-$           & $-$    
	& 4.241	& 36.26	& 4.233	& 9.457	& $-1.0\times 10^{-6}$
\\
(c2)& const.	& 3.75  & $-$           & $-$      
	& 4.882	& 11.57	& 4.836	& 7.720	& $-1.0\times 10^{-5}$
\\
(c3)& const.	& 3.70  & $-$           & $-$      
	& 5.538	& 8.227	& 5.266	& 6.921	& $-1.0\times 10^{-4}$
\\
(l1)& linear   	& $-$   & $0.03,\; 3.70$& $-$      
	& 5.115	& 46.71	& 4.444	& 10.13	& $-1.0\times 10^{-4}$
\\
(l2)& linear	& $-$   & $0.01,\; 3.70$& $-$      
	& 5.224 & 12.74 & 4.845 & 8.087 & $-1.0\times 10^{-5}$
\\
(pl1)& power-law	& 3.70  & $-$   & 0.02      
	& 4.872	& 20.97	& 4.520	& 9.108	& $-1.0\times 10^{-5}$	
\\
(pl2)& power-law	& 3.60  & $-$   & 0.05      
	& 5.065	& 61.98	& 4.312	& 11.06	& $-1.0\times 10^{-5}$
\\
(pl3)& power-law  	& 3.66  & $-$   & 0.02 
	& 5.157 & 15.46 & 4.694 & 8.596 & $-1.0\times 10^{-5}$
\\
(pl4)& power-law  	& 3.68  & $-$   & 0.03
	& 4.854 & 32.13 & 4.395 & 9.889 & $-1.0\times 10^{-6}$
\\ ~ \\\hline
\end{tabular}
\caption{ \small{ Main properties of the two non constant angular
momentum models considered. From left to right the columns report: the
type of angular momentum distribution (see Section \ref{GRT_ncamd} for a
description of $~{\tilde \ell}_{\rm c}$, $\alpha$, $\beta$ and $q$), the
inner and the outer radii of the torus ${\widetilde \varpi}_{\rm in}$ and
${\widetilde \varpi}_{\rm out}$, the radial position of the maximum
density in the torus ${\widetilde \varpi}_{\rm max}$, and the value of
the potential gap chosen at the inner edge of the torus. }}
\end{center}
\end{minipage}
\end{table*}

	Finally, it should be noted that, as for the Newtonian case,
expression (\ref{gr_epicyclic}) predicts a {\it zero epicyclic frequency
for orbital motion with constant specific angular momentum.} Note that
this holds true only when the specific angular momentum is defined as
$\ell\equiv -u_\phi/u_t$, in which case
\begin{equation}
\label{vanish_k} 
\frac{d\ell}{d\varpi}=\frac{d}{d\varpi}(\varpi^2 e^{-2\nu}\Omega)
	=\varpi e^{-2\nu}\left(2\Omega +\varpi\frac{d \Omega}{d \varpi}
	-2\varpi\Omega\frac{d \nu}{d \varpi}\right) \ . 
\end{equation} 

	To the best of our knowledge, this result has not been discussed
before in the literature for finite-size fluid
configurations. Furthermore, it is relevant for the physical
interpretation of the modes found in the numerical calculations of
Zanotti et al. (2003) as it indicates that the oscillations found in
those simulations cannot be associated (at least within the validity of a
vertically integrated approach) with epicyclic oscillations, but should
be associated with other restoring forces, most notably pressure
gradients.

\section{Perturbations of Relativistic Tori: a Global Analysis}
\label{GRT_ga}

	A global analysis of the axisymmetric oscillations modes of a
relativistic torus consists of solving the system of equations
(\ref{euler-r})--(\ref{cont_gr}) as an eigenvalue problem, treating the
perturbed quantities as eigenfunctions and the unknown $\sigma$ as the
eigenfrequency (see Rodriguez et al., 2002 for the solution of the
equivalent problem for relativistic thin discs).

	It is worth remarking that equation (\ref{bernoulli}) basically
states that a condition for the existence of stationary, geometrically
thick fluid configurations orbiting around a Schwarzschild black hole is
that the distribution of angular momentum is a non-Keplerian one. As long
as this condition is met, in fact, the left-hand-side of equation
(\ref{bernoulli}) will not be zero and the pressure gradients will expand
the disc in the vertical direction. Of course, different distributions of
angular momentum will define tori with different properties such as:
different inner and outer radii, different positions for the maximum in
density, etc. Among the infinite number of possible angular momentum
distributions, those that produce tori with a finite radial extension are
of particular interest, since they allow for the existence of globally
trapped, axisymmetric modes of oscillation.

	Because of the degeneracy in the functional form for
$\ell(\varpi)$, we have constructed several different models that we have
reported in Table~1 and that we will be discussing in detail in the
following Sections. This approach to the problem is clearly more
expensive as the eigenvalue problem needs to be solved for a large number
of models; however it also allows one to investigate how the axisymmetric
oscillations depend on the angular momentum distribution. When features
are found which do not depend on a specific choice for the distribution
of angular momentum and that are thus indication of a universal
behaviour, this approach can be very rewarding as we will discuss in the
following Sections.

	However, before discussing the results of the numerical solution
of the eigenvalue problems, it is useful to review briefly the numerical
approach followed in the solution of the eigenvalue problem and the
boundary conditions imposed.

\subsection{Numerical method}

\label{nm}

	Equations (\ref{euler-r})--(\ref{cont_gr}), supplemented with
suitable boundary conditions at both edges of the numerical grid,
represent a standard two-point boundary value problem, for which a
``shooting'' method can be used. In practice, given a trial value for the
eigenfrequency $\sigma$, two solutions for the unknown quantity (either
$\delta Q$ or $\delta U$) are found starting from the two edges of the
numerical grid at the inner, ${\varpi}_{\rm in}$, and outer,
${\varpi}_{\rm out}$, radii of the torus. These two solutions are then
matched at an arbitrary point ${\varpi}_{_{\rm M}}$ in the domain, where
the Wronskian of the left and right solutions is evaluated. This
procedure is iterated until a zero of the Wronskian is found with the
desired accuracy, thus determining both the eigenfrequency and the
eigenfunctions of the mode considered.

	Note that because of the linearity of the present analysis, the
solutions are invariant after a rescaling of all the quantities by a
constant. In practice, this freedom allows to choose the initial guess
for one of the eigenfunctions at one edge of the numerical grid to be a
freely specifiable number, which we typically set to one. Furthermore,
since the spacetime is uniquely described once the mass $M$ of the black
hole has been fixed, we can use it to rescale all of the relevant
quantites involved in the calculations. The new dimensionless quantities,
indicated with a ``tilde'', are then defined as

\begin{eqnarray}
\varpi&  \equiv& \widetilde{\varpi}\left(\frac{M}{M_\odot}\right) 
	\  \left(\frac{GM_\odot}{c^2}\right)\ , 
\\ \nonumber \\
\Sigma &\equiv &\widetilde{\Sigma}\left(\frac{M}{M_\odot}\right)^{-1} 
		\  \left(\frac{GM_\odot}{c^2}\right)^{-2}\ M_\odot\ , 
\\ \nonumber \\
P&\equiv&\widetilde{P}\left(\frac{M}{M_\odot}\right)^{-1} \
		\left(\frac{GM_\odot}{c^2}\right)^{-2}\ M_\odot \ c^2\ , 
\\ \nonumber \\
K& \equiv& \widetilde{K} \left(\frac{M}{M_\odot}\right)^{\Gamma-1}\ 
	\left(\frac{GM_\odot}{c^2}\right)^{2(\Gamma-1)}
	\ (M_\odot)^{1-\Gamma}\ c^2\ , 
\\ \nonumber \\
\sigma &\equiv &\tilde{\sigma}\left(\frac{M}{M_\odot}\right)^{-1}\ 
		\left(\frac{GM_\odot}{c^2}\right)^{-1}\   c\ , 
\\ \nonumber \\
\ell& \equiv& \tilde{\ell} \left(\frac{M}{M_\odot}\right)\  
	\left(\frac{GM_\odot}{c^2}\right) \  c \ .
\end{eqnarray}
Unless specified differently, all of the results presented hereafter will
be given in terms of these dimensionless quantities.

\subsection{Boundary conditions}
\label{bcs}

	Suitable boundary conditions need to be specified at the inner
and outer edges of the vertically integrated torus for the solution of
the set of equations (\ref{euler-r})--(\ref{cont_gr}). As for
oscillations in stars, we here assume the perturbed surface of the torus
to be where the Lagrangian pressure perturbation vanishes, i.e. where
\begin{equation}
\label{lpeq0}
\Delta p = 0\ .
\end{equation}
Because the Eulerian and Lagrangian descriptions are linked through the
relation
\begin{equation}
\label{eul_lag}
\Delta = \delta + {\cal L}_{\vec \xi}  \ ,
\end{equation}
where ${\cal L}_{\vec \xi}$ is the Lie derivative along the Lagrangian
displacement three-vector $\vec \xi$, condition (\ref{lpeq0}) can also be
written as 
\begin{equation}
	\delta p + \xi^j\partial_j p = 0 \ , 
\end{equation}
from which it follows that $\Delta p=0=\delta p$ for a polytropic fluid
configuration whose rest-mass density vanishes at the surface (Tassoul,
1978). Note also that while in the Cowling approximation the Eulerian
perturbation of the metric, $\delta g_{ab}$ is zero, the Lagrangian
perturbation $\Delta g_{ab} = \delta g_{ab} + {\cal L}_{\vec \xi}g_{ab} =
\nabla_a\xi_b + \nabla_a\xi_b$ is, in general, nonzero.

	Using the relation between the perturbed 4-velocity $u^\alpha$
and the Lagrangian perturbation  of the metric (see Friedman 1978),
\begin{equation}
\Delta u^\alpha = \frac{1}{2}u^\alpha u^\beta u^\gamma
		\Delta g_{\beta\gamma}\ ,
\end{equation}
which, in our case, reduces to the relation,
\begin{equation}
\xi^\varpi = \frac{ie^{\nu-\lambda}\delta V^{\hat{\varpi}}} {\sigma}\ ,
%
\end{equation}	
we can write the boundary condition (\ref{lpeq0}) simply as
\begin{equation}
\label{boundary}
\delta Q + \frac{e^{\nu-\lambda}}{\sigma(E+P)}
		\frac{d P}{d \varpi}\delta U
			 = 0 \ .
\end{equation}
Expression (\ref{boundary}), taken as a linear relation between $\delta
Q$ and $\delta U$, provides us with the boundary conditions to be imposed
at the inner and outer edges of the torus.

\section{Constant Specific Angular Momentum Tori}
\label{num_1}

	The simplest and most studied choice for the distribution of
angular momentum is the one in which $\ell=\ell_{\rm c}={\rm const.}$,
with $\ell_{\rm c}$ being chosen in the interval between the specific
angular momentum at the marginally stable orbit $\ell_{\rm ms} =
3\sqrt{6}/2 \simeq 3.67$ and the one at the marginally bound orbit
$\ell_{\rm mb} = 4$, to yield a disc of finite size. Besides yielding a
simpler approach, constant specific angular momentum tori benefit from
having a radial epicyclic frequency which is identically zero
[cf. eqs. (\ref{gr_epicyclic}) and (\ref{vanish_k})], thus leaving
pressure gradients as the only possible restoring forces. As a result, a
constant angular momentum distribution provides an interesting limit for
the properties of $p$-mode oscillations that will not be influenced by
centrifugal effects.

	The procedure for the construction of the background model for
the torus can be summarized as follows. Firstly, after a value $\ell_{\rm
c}$ for the constant specific angular momentum has been chosen , the
locations of the cusp $\varpi_{\rm cusp}$ and of the maximum density in the
torus\footnote{Note that the position of the maximum density in the torus
is traditionally referred to as the ``centre'' of the torus, although its
coordinate centre is, of course, at $\varpi=z=0$.}  $\varpi_{\rm max}$
are determined as the positions at which the Keplerian specific angular
momentum is equal to $\ell_{\rm c}$. The inner and outer edges of the
torus can then be calculated after requiring that they lie on the same
equipotential surface, i.e. 
\begin{equation}
u_t (\varpi_{\rm in}) = u_t (\varpi_{\rm cusp})
\exp(\Delta W_{\rm in}) = u_t (\varpi_{\rm out}) \ ,
\end{equation}
where $\Delta W_{\rm in}$ is the value of the potential gap chosen at the
inner edge of the torus and is therefore a measure of how much the torus
``fills'' its Roche lobe. Clearly, the case in which $\Delta W_{\rm
in}=0$ corresponds to a torus filling its outermost closed equipotential
surface (i.e. the one possessing a cusp)

\begin{figure}
\centerline{
\psfig{file=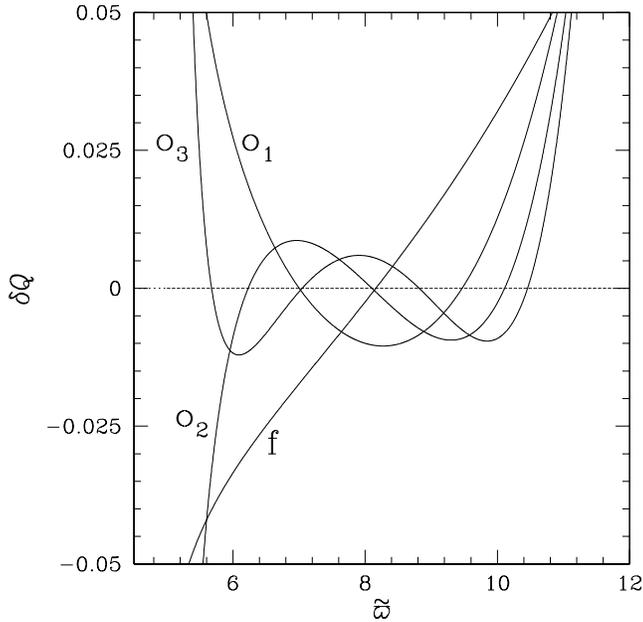,width=8.75cm,angle=0}
	}
\caption{ \small{ Eigenfunctions for $\delta Q=\delta P/(E+P)$ as a
function of the radial coordinate for a constant angular momentum
disc. Only the fundamental mode $f$ and the first three overtones,
denoted by $o_1$, $o_2$ and $o_3$, have been reported. The data refers to
model (c2) of Table~1 and the units on the vertical axis are
arbitrary.}}
\label{Q_l0}
\end{figure}

	Having determined $\varpi_{\rm in}$, it is then possible to
calculate the angular velocity distribution as
\begin{equation}
\Omega = \Omega_{\rm c} \left(\frac{\varpi - 2M}{\varpi_{\rm c} - 2M}\right)
	\left(\frac{\varpi_{\rm c}}{\varpi}\right)^3
	\ ,
\end{equation}
where $\Omega_{\rm c}$ is angular velocity at the radial position of cusp
$\varpi_{\rm cusp}$. Finally, after making a choice for the parameters of
the equation of state (and therefore for $k$ and $\gamma$) the equation
for the hydrostatic equilibrium (\ref{vert_eq}) can be integrated, either
numerically or analytically (Kozlowski et al., 1978), to build the torus
model which will serve as background for the introduction of the
perturbations.

	When the specific angular momentum is constant within the torus,
also the perturbation equations become simpler and, in particular, the
second term in (\ref{euler-phi}) vanishes exactly [cf. equation
(\ref{vanish_k})] and the resulting set of equations can then be written
as
\begin{eqnarray}
\label{euler-rr}
&&  \sigma \delta U + A e^{-\nu-\lambda} \frac{d (\delta Q)}{d \varpi} 
	= 0 \ , 
\\ \nonumber	\\
\label{euler-phi_bc}
&&      \delta V^{\hat{\phi}} + 
	A\Omega e^{-2\nu} \varpi \delta Q = 0 \ ,
\\\nonumber\\
\label{baryon}
&&  \sigma \delta Q + e^{\nu-\lambda}\frac{\Gamma P}{E+P}
	\frac{d (\delta U)}{d \varpi} + 
	\frac{e^{\nu-\lambda}\Gamma P}{E+P}
	\times
\nonumber \\
&& \hskip 1.0 truecm 
	\times\left[
	2\frac{d \nu}{d \varpi} +
	\frac{1}{\varpi}-\frac{1}{A}\frac{d A}{d \varpi}
	+ \frac{1}{\Gamma P}\frac{d P}{d \varpi}\right] \delta U = 0 \ . 
\end{eqnarray}

	Equations (\ref{euler-rr})--(\ref{baryon}) have been solved
numerically for a number of different models, whose main properties, such
as the inner and outer radii ${\widetilde \varpi}_{\rm in}, {\widetilde
\varpi}_{\rm out}$, the position of the maximum in the surface density
distribution ${\widetilde \varpi}_{\rm max}$, and the value of the
specific angular momentum at the inner edge of the disc $\ell_{\rm c}
\equiv \ell(\varpi_{\rm in})$, are summarized in Table~1. As a
representative example, we present in Figures~\ref{Q_l0} and 2 the
results for a torus model with $\ell_{\rm c} = 3.75$ [i.e. model (c2) of
Table~1].

\begin{figure}
\centerline{
\psfig{file=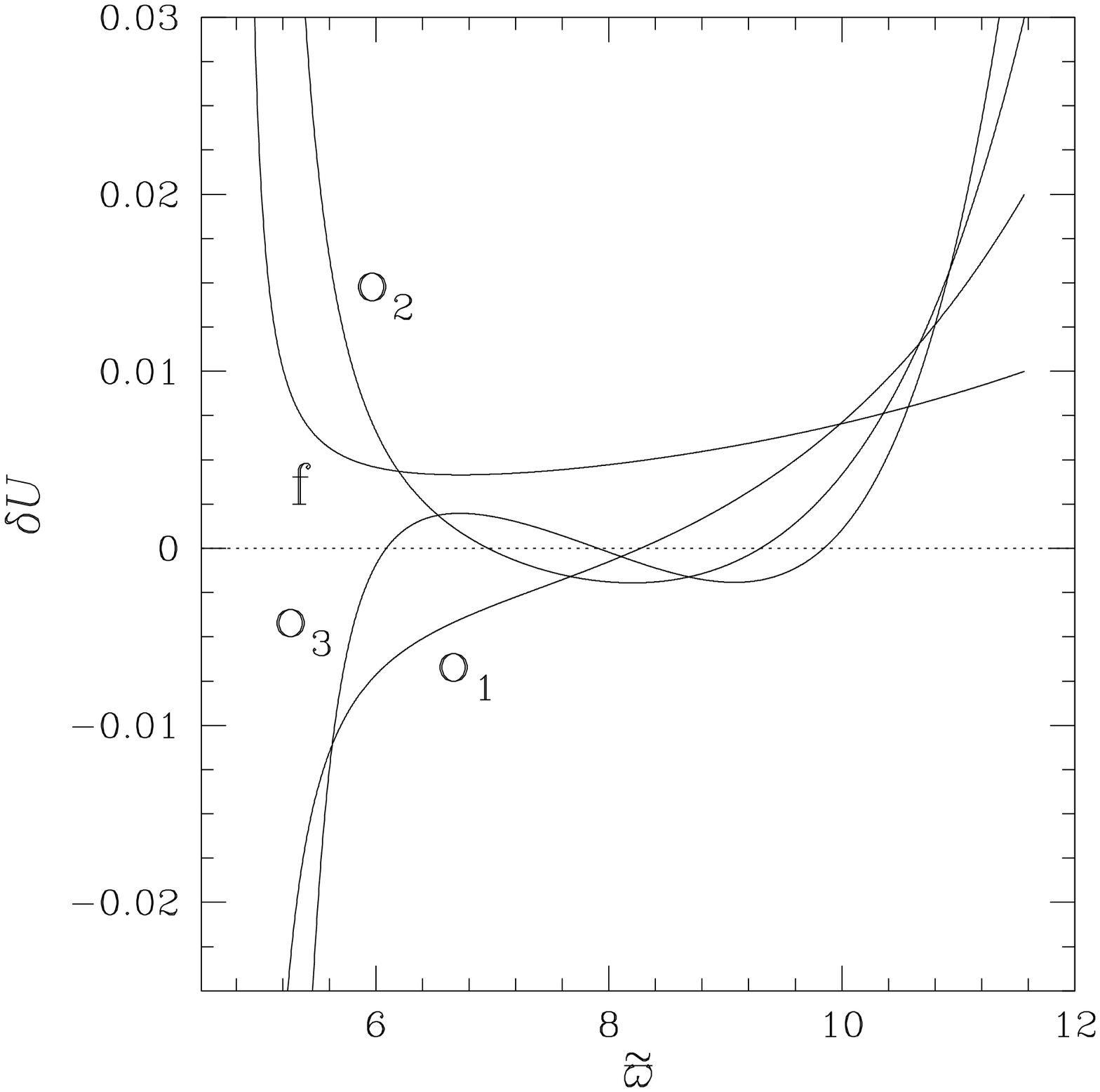,width=8.75cm,angle=0}
	}
\caption{ \small{Eigenfunctions for $\delta U=i\delta V^{\hat \varpi}$ as
a function of the radial coordinate for a constant angular momentum
disc. Only the fundamental mode $f$ and the first three overtones,
denoted by $o_1$, $o_2$ and $o_3$, have been reported. The data refers to
model (c2) of Table~1 and the units on the vertical axis are arbitrary.}}
\label{U_l0}
\end{figure}

	Figure~\ref{Q_l0}, in particular, shows the solution for the
eigenfunction $\delta Q$ for both the fundamental mode $f$ and for the
first three overtones $o_1, o_2$ and $o_3$. Similarly, Figure~\ref{U_l0}
shows the fundamental mode and the first three overtones for the
eigenfunction $\delta U$, which has always one node less than $\delta
Q$. Note that the choice of a very small potential barrier introduces
complications in the numerical solution of the eigenvalue problem. For a
model with $\Delta W_{{\rm in}}=0$, in fact, the fluid elements at the
inner edge are just marginally stable to accretion onto the black
hole. This behaviour is reflected in the eigenfunctions $\delta Q$ and
$\delta U$, which tend to diverge at the inner and outer edges of the
torus. To avoid this problem (still partially visible in
Figures~\ref{Q_l0} and \ref{U_l0}) we have used a small but nonzero
value, $\Delta W_{{\rm in}}=-10^{-5}$, for the potential barrier.

	A careful look at Figures~\ref{Q_l0} and \ref{U_l0} reveals that
the modes $f$ and $o_2$ have similar properties at the inner edge;
furthermore, during each half-period, they have signs which are opposite
to those of the modes $o_1$ and $o_3$. This is an important feature
indicating that not all modes will behave in the same manner at the inner
edge where mass-loss and accretion onto the black hole take place. While
a conclusion on the this behaviour cannot be drawn on the basis of the
present linear analysis, in which the total mass is conserved exactly
(${\widetilde \Sigma} \delta U=0$ at the inner and outer edges for all
modes), the functional form of the eigenfunctions for $\delta Q$ and
$\delta U$ seems to suggest that mass accretion through the cusp will be
possible preferably at the frequencies corresponding to the $f$, and
$o_{2n}$ modes (see also the discussion below for a comparison with the
numerical calculations presented in Zanotti et al., 2003).

	The eigenfrequencies corresponding to model (c2), as well as for
all of the other models in Table~1, are listed in Table~2, where we have
included the frequencies up to the fourth overtone. A careful look at
Table~2 reveals that the computed eigenfrequencies are, at least for the
first few modes, in a sequence 2:3:4:$\ldots$, to a good precision. As
the order of the mode increases, this simple sequence is no longer
followed and a different one appears. The fact that the lowest order
modes appear at frequencies that are in small integer ratios is not
particularly surprising if $p$ modes are to be interpreted as sound waves
trapped within the cavity represented by the confined torus. In this
case, in fact, one would indeed expect that modes should be trapped and
with wavelengths that are multiples of the trapping lengthscale,
i.e. $\lambda_n = [2/(2+n)] L$, where $n=0,1,\ldots$. Stated differently,
the results reported in Table~2 are consistent with the idea that $p$
modes can be associated with sound waves trapped in the torus and having
frequencies which are multiples of the fundamental one and given by
$\sigma_n = c_s/\lambda_n = [(2+n)/2] f$.

	Note that while these $p$ modes follow the same definition and are
similar in nature to the ones discussed by Nowak and Wagoner (1991, 1992)
for thin discs, their trapping is not produced by a turning of the
relativistic radial epicyclic frequency at smaller radii. Rather, the
relativistic radial epicyclic frequency in discs in which pressure
gradients play an important role, has been found to be monotonically
decreasing for all of the disc models considered here [cf. equation
(\ref{gr_epicyclic}], Fig.~\ref{epic_1}]. An important consequence of
this is that the $p$ modes for thick discs are {\it not restricted} to be
trapped in special parts of the disc, as it happens for the corresponding
$p$ modes for thin discs (Nowak \& Wagoner 1991, 1992). In particular,
they need to be present only in the (rather small) inner regions of the
disc, where they can produce only a slight modulation in the disc
emission. Similarly, they are not constrained to be trapped the outer
regions of the disc, where the location of the outer radius remains
uncertain and the corresponding frequencies very small. On the contrary,
the modes discussed here are present over the whole torus, which then
behaves as a {\it single} trapping cavity. This is a substantial
difference, giving these modes the possibility of possessing frequencies
comparable with observations and of causing non-negligible modulations in
the disc emission.

\begin{table}
\label{tab2}
\begin{center}
\begin{tabular}{l|ccccccc}\hline
Model & $f$ & $o_1$ & $o_2$ & $o_3$ & $o_4$ \\
 \hline 				\\
(c1)  & 0.00838 & 0.01227 & 0.01604 & 0.01973 & 0.02337	\\
(c2)  & 0.02017 & 0.02950 & 0.03700 & 0.04322 & 0.04869	\\
(c3)  & 0.01898 & 0.02764 & 0.03514 & 0.04230 & 0.04933	\\
(l1)  & 0.00602 & 0.00831 & 0.01060 & 0.01287 & 0.01514	\\
(l2)  & 0.01992 & 0.02856 & 0.03618 & 0.04330 & 0.05019 \\
(pl1) & 0.01447 & 0.02114 & 0.02733 & 0.03324 & 0.03894	\\
(pl2) & 0.00426 & 0.00616 & 0.00803 & 0.00988 & 0.01172	\\
(pl3) & 0.01810 & 0.02635 & 0.03764 & 0.04073 & 0.04742	\\
(pl4) & 0.00948 & 0.01379 & 0.01792 & 0.02196 & 0.02593	\\
 ~ 			\\
\hline 
\end{tabular}
\caption{ \small{Eigenfrequencies of the fundamental $f$ mode and of the
first four overtones $o_n$ of the vertically integrated models described
in Table~1. All the frequencies are given in normalized units.}}
\label{freq-lin}
\end{center} 
\end{table}

\begin{figure}
\centerline{
\psfig{file=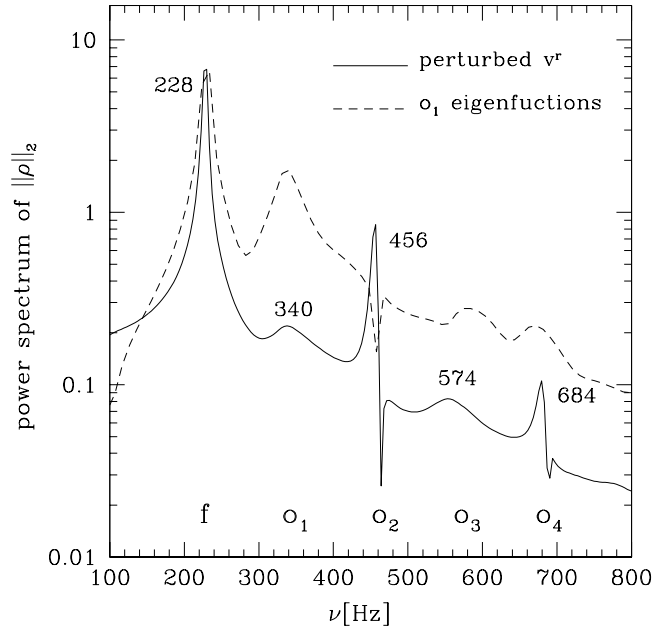,angle=0,width=8.75cm}
        }
\caption{ \small{Power spectra of the $L_2$ norm of the rest-mass density
for the torus model (c2) in the Table~1 of Zanotti et al.
(2003). Different lines refer to models with different initial
perturbations. The solid line, in particular, refers to initial data in
which a global perturbation has been introduced in the radial velocity
field only [cf. equation (15) of Zanotti et al. (2003)]. The dotted
line, on the other hand, refers to initial data in which a perturbation
based on the computed eigenfunctions of the $o_1$ mode has been
introduced both in the radial velocity and in the density. The two
spectra have been rescaled in order to match in the power of the
fundamental frequency.}}
\label{pspectra}
\end{figure}

	Having solved the eigenvalue problem for $p$ modes in vertically
integrated relativistic tori with constant specific angular momentum, we
can now proceed to a comparison with the fully nonlinear two-dimensional
simulations discussed by Zanotti et al. (2003). This is presented in
Figure~\ref{pspectra} where we have plotted the power spectra of the
$L_2$ norms of the rest-mass density for the torus model (c2) in
Table~1. The solid line, in particular, refers to initial data in which a
global perturbation has been introduced in terms of the radial velocity
field of a relativistic spherical accretion solution [cf. equation (15)
of Zanotti et al., 2003].  Clearly, the power spectrum for the continuous
line shows the presence of peaks appearing in a simple small integer
sequence 2:3:4:$\ldots$, providing convincing evidence that the
oscillations triggered in the simulations of Zanotti et al. (2003)
correspond indeed to $p$-modes oscillations of perturbed relativistic
tori. It should also be noted that this sequence is not the same shown by
the peaks in the power spectrum of the mass accretion rate onto the black
hole, which instead are only in the sequence 1:2:3:$\ldots$ (cf. Figure~7
of Zanotti et al., 2003).  This difference remains puzzling, but could be
explained on the basis of the behaviour of the eigenfunctions for the $f$
and $o_{2n}$ modes at the inner edge discussed above. In this case, in
fact, all of the $o_{2n+1}$ modes would have vanishingly small
perturbations at the inner edge of the disc and would not produce a mass
accretion at those frequencies, as shown by Figure~7 of Zanotti et al.
(2003).

	The knowledge of the eigenfunctions of $p$ modes in relativistic
tori can also be used to go a step beyond a simple comparison with
numerical simulations and actually use the latter to perform
investigations of ``numerical'' discoseismology. In other words, having a
complete picture of the oscillation properties and an accurate
two-dimensional numerical code to simulate them, it is possible to
investigate the dynamical response of a relativistic thick disc as a
result of the introduction of suitably selected perturbations. As a
concrete example, we report with the dashed line in Figure~\ref{pspectra}
the power spectrum of the $L_2$ norm of the rest-mass density for the
torus model (c2) in Table~1. The important difference with the
corresponding spectrum indicated with a solid line is that the dashed
line refers to a simulation having as initial data perturbations based on
the computed eigenfunctions for $\delta U$ and $\delta Q$ of the $o_1$
mode. This represents a selective excitation of the $o_1$ mode and, as a
result, the corresponding power in the first overtone is increased by
almost a factor of ten (the power in the $o_2$ mode is instead decreased
of a similar amount). This behaviour further confirms that the $p$ modes
derived in this perturbative analysis correspond to the modes simulated
numerically by Zanotti et al. (2003). Finally, it should be noted that
the power spectrum corresponding to initial data containing only an
$o_1$-mode perturbation shows that also the other modes have been
excited, most notably the fundamental one. This is due partly to a
mode-mode coupling which transfers energy from one mode to the other
ones, but also to the error introduced using eigenfunctions derived for a
vertically integrated model, and that cannot reproduce the corresponding
eigenfunctions of a two-dimensional disc exactly.

\begin{figure}
\centerline{
\psfig{file=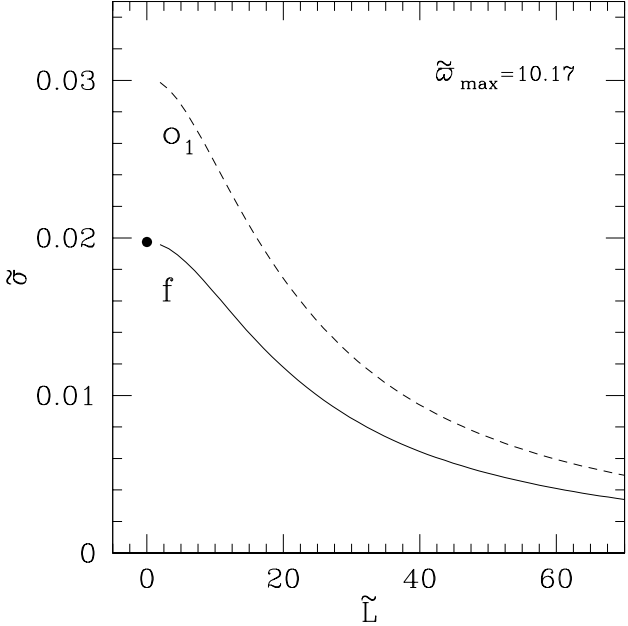,angle=0,width=8.75cm}
        }
\caption{\small{Eigenfrequencies for the fundamental $f$ (solid line) and
for the first overtone $o_1$ (dashed line) of axisymmetric $p$ modes for
$\ell=\;$const. tori. The two lines refer to sequences of tori having the
same radial position for the ``centre'' ${\widetilde \varpi}_{\rm max} =
10.17$, but different radial extensions ${\widetilde L} \equiv
{\widetilde \varpi}_{\rm out} - {\widetilde \varpi}_{\rm in}$. The filled
dot represents the radial epicyclic frequency for a circular orbit at
${\widetilde \varpi}_{\rm max}$ and thus the value at which the
fundamental $p$-mode frequency tends in the limit of a vanishing torus
size.}}
\label{f_vs_L}
\end{figure}

	There is a final aspect of the axisymmetric $p$ modes discussed
so far which is worth underlining. This is illustrated in
Figure~\ref{f_vs_L} where we have plotted the values of the
eigenfrequencies for the fundamental mode $f$ (solid line) and for the
first overtone $o_1$ (dashed line) for large number of tori. All of the
points on the solid and dashed curves in Figure~\ref{f_vs_L} represent
the solution of the eigenvalue problem and the two sequences have been
calculated for tori with fixed ``centre'' ${\tilde \varpi}_{\rm max} =
10.17$, but having different radial extensions ${\widetilde L} \equiv
{\widetilde \varpi}_{\rm out} - {\widetilde \varpi}_{\rm in}$. As one
would expect for modes behaving effectively as sound waves, the
eigenfrequencies decrease as the radial extent of the torus
increases. Furthermore, the frequencies for $f$ and $o_1$ maintain a
harmonic ratio 2:3 for all the values of ${\tilde L}$. Most importantly,
however, the filled dot shown in Figure~\ref{f_vs_L} for ${\tilde L}=0$
represents the radial epicyclic frequency for a circular orbit at
${\tilde \varpi}_{\rm c}$, i.e. ${\tilde \sigma} = \sqrt{(1 - 6
/{\widetilde \varpi}_{\rm c})GM/\varpi^3_{\rm c}}$.

	This is an important result that provides a simple interpretation
of the nature of the axisymmetric $p$ modes of thick discs: {\it the
radial epicyclic frequency at the position of maximum density represents
the value at which the fundamental $p$-mode frequency tends in the limit
of a vanishing torus size}. The importance of this result is that while
different models for the tori, using, for instance, different equations
of state or different angular momentum distributions, will produce
different slopes for the solid and dashed lines in Figure~\ref{f_vs_L},
all of the sequences will terminate on the filled dot when $L\rightarrow
0$, that is, when finite-size effects will cease to be relevant and each
torus will effectively behave as a ring of particle in circular
orbits. Because this occurs only in the limit $L\rightarrow 0$, it should
not surprise that it is valid also for sequences of tori with constant
distributions of specific angular momentum for which, as discussed in
Section~\ref{GRT_la}, the epicyclic frequency is zero. Indeed, this
result is totally independent of the distribution of specific angular
momentum and will be confirmed in the following Section, where tori with
non-constant specific angular momentum distributions are considered.  An
analytic proof of this conclusion is provided in Appendix A.

\section{Non-constant Specific Angular Momentum Tori}
\label{GRT_ncamd}

	While the study of constant angular momentum discs offers several
advantages and simplifies the equations, realistic disc are likely to
have angular momentum distributions that are not constant.  For this
reason, and in order to assess the validity of the results derived so far
with more generic distributions of angular momentum, we have extended our
mode analysis also to discs with non-constant angular momentum. The first
step in this direction consists of determining which distribution of
specific angular momentum $\ell = \ell(\varpi)$ should be specified in
the construction of an equilibrium model. This choice is, to some extent,
arbitrary with the only constraint being given by {\it Rayleigh's
criterion} for the dynamical stability against axisymmetric perturbations
(Tassoul 1978). This condition, however, is not very strong and simply
requires that $d\ell/d\varpi \geq 0$, so that even a constant angular
momentum distribution is stable, although only marginally.

	Some guidance in this choice can come from numerical simulations
and indeed calculations of torus formation performed by Davies et
al. (1994) with smooth particle hydrodynamics (SPH) techniques have shown
that the final configuration consists of a core object surrounded by a
torus whose angular momentum distribution is close to a power-law in
radius, with index $0.2$. In our computations we have therefore adopted
two simple and different distributions for the specific angular momentum
that can be expressed analytically. We refer to the first one as the {\it
``linear''} distribution, in which
\begin{equation}
\ell = \alpha \varpi + \beta \ ,
\end{equation}
where $\alpha$ and $\beta$ are adjustable constant coefficients.
Similarly, we refer to the second one as the {\it ``power-law''}
distribution, in which instead
\begin{equation}
\ell = \ell_{\rm c} \varpi^q\ ,
\end{equation}
where $q$ is also an arbitrary parameter. The values chosen for $\alpha,
\;\beta, \;\tilde{\ell}_{\rm c}$ and $q$ for the representative models
discussed here have been summarized in Table~1. Note that we have
deliberately chosen small values for both $\alpha$ and $q$ and this is to
avoid the construction of unpertubed models that differ significantly in
radial extension from the ones built with a constant specific angular
momentum distribution\footnote{It is not difficult to realize that a
constant specific angular momentum distribution yields the most compact
tori among those having the same angular momentum at the cusp.}.

	The construction of the equilibrium models proceeds in this case
as discussed in the previous Section, namely, by first fixing an initial
value for the specific angular momentum at the cusp, then by calculating
the radial extension of the torus in terms of the potential barrier
$\Delta W_{\rm in}$, and finally by integrating numerically equation
(\ref{quasi}) for the chosen distribution of $\Omega(\varpi)$.

	The full system of the perturbed equations
(\ref{euler-r})--(\ref{cont_gr}), can now be rewritten in a more compact
form as
\begin{eqnarray}
&& 0= \sigma \delta Q + e^{\nu-\lambda}\frac{\Gamma P}{E +P}
	\frac{d (\delta U)}{d\varpi} + 
\nonumber \\
&& \hskip 0.5truecm
	\frac{e^{\nu-\lambda}\Gamma P}{E +P}
	\left[2\frac{d \nu}{d \varpi} + 
	\frac{1}{\varpi}-\frac{1}{A}\frac{d A}{d \varpi}
	+\frac{1}{\Gamma P}\frac{d P}{d \varpi} \right] \delta U -
\nonumber\\
&& \hskip 4.0 truecm
	\sigma\frac{e^\nu \Omega}{A}
	\left(\frac{\varpi}{E+P}\right)
	\frac{d P}{d \varpi} \ ,
\\ \nonumber \\
\label{baryon2}
&& 0= \sigma \delta U + A e^{-\nu-\lambda} 
	\frac{d (\delta Q)}{d\varpi} + 
\nonumber\\
&& \hskip 2.5 truecm
	2e^{-\lambda}\Omega
	\left(1+
	\frac{\varpi}{E+P}\frac{d P}{d \varpi}\right) \delta W \ ,
\\ \nonumber \\
\label{euler-r2}
&&  0= \sigma \delta W +
	\left(\frac{d \Omega}{d \varpi} + 
	\frac{2}{\varpi}\Omega - 
	2 \Omega \frac{d \nu}{d \varpi} \right)\varpi
	e^{-\lambda} \delta U\ ,
\end{eqnarray}
and can be solved adopting the same numerical technique used for the
constant angular momentum models. 

\begin{figure}
\centerline{
     \psfig{file=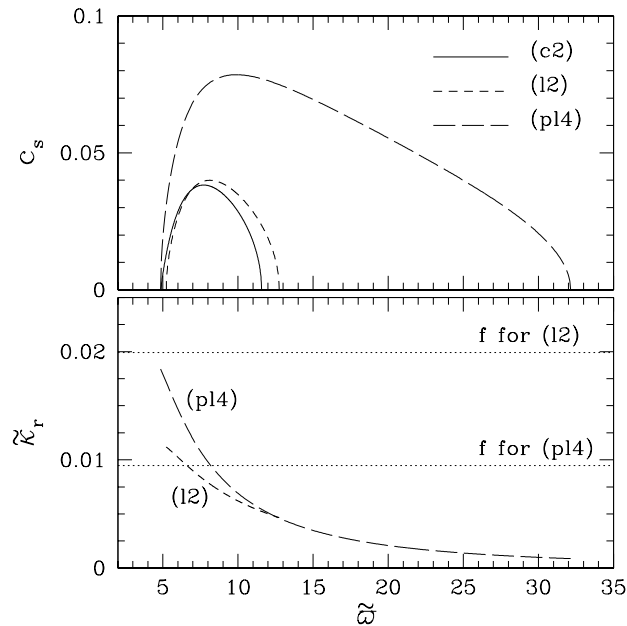,width=8.75cm,angle=0}
}
\caption{ \small{ The upper panel shows the relativistic sound velocity
for representative tori models having either a constant [solid line;
model (c2)], a linear [short-dashed line; model (l2)] or a power-law
[long-dashed line; model (pl4)] distribution of specific angular
momentum. The lower panel, instead, shows the dimensionless radial
epicyclic frequency for the three models shown in the upper panel, with
the horizontal lines showing the fundamental frequencies for models (l2)
and (pl4). All quantities are expressed in normalized units. }}
\label{epic_1}
\end{figure}

	The upper panel of Figure~\ref{epic_1} shows the relativistic
sound velocity for representative unperturbed tori models having either a
constant (solid line), a linear (dotted line) or a power-law (dashed
line) distribution of specific angular momentum [The data refers to
models (c2), (l2) and (pl4) of Table~1, respectively.]. Note that while
all of the curves refer to models with a polytropic EOS, each curve does
not depend on the value chosen for the polytropic constant $k$ and hence
on the mass of the disc (see Appendix B for a proof of this). The lower
panel of Figure~\ref{epic_1}, on the other hand, shows the radial
epicyclic frequency $\kappa_{\rm r}$ as calculated from expression
(\ref{gr_epicyclic}) for the three models shown in the upper panel (we
recall that $\kappa_{\rm r}=0$ when $\ell={\rm const.}$). The horizontal
lines in the lower panel refer to the fundamental frequencies for models
(l2) and (pl4) and their relative position with respect to the curves for
$\kappa_{\rm r}$ will be important for the appearence of an
evanescent-wave region in the inner parts of the torus (see the
discussion below).

	We have summarized in the different panels of Figure~\ref{ncsamd}
the eigenfunctions for $\delta Q$, $\delta U$ and $\delta W$ for tori
with a linear [panels (a)--(c)] and a power-law [panels (d)--(f)]
distribution of specific angular momentum. The eigenfunctions refer, in
particular, to models (l1) and (pl3) of Table~1 and are shown in the
fundamental as well in the first three overtones. Similarly, are
summarized in Table~2 also the eigenfrequencies computed for a number of
tori with non-constant specific angular momentum.

\begin{figure}
\centerline{
     \psfig{file=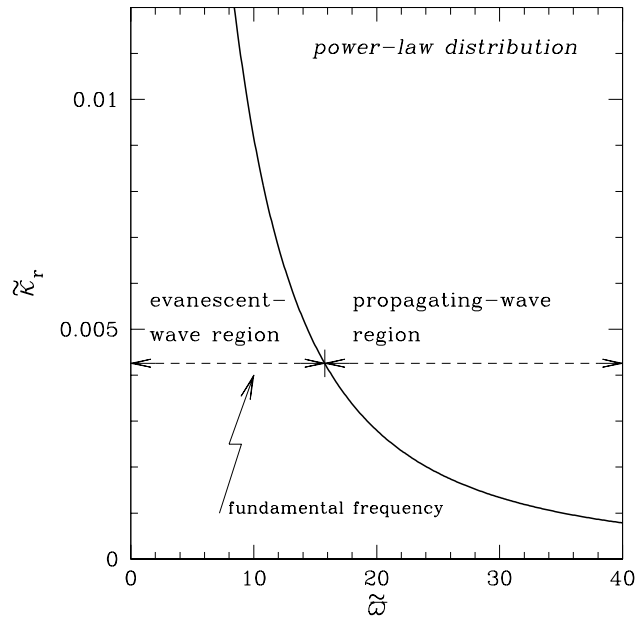,width=8.75cm,angle=0}
	}
\caption{ \small{Schematic propagation diagram for a thick disc. The
thick solid line shows the values of the relativistic radial epicyclic
frequency ${\tilde \kappa}_{\rm r}$ in the inner regions of the disc,
while the horizontal dashed line the value of the fundamental frequency
$f$. The data is presented in normalized units and refers to model (pl2)
of Table~1. See the main text for a discussion.}}
\label{evan}
\end{figure}

\begin{figure*}
\begin{center}
\hspace{0.001cm}
\psfig{file=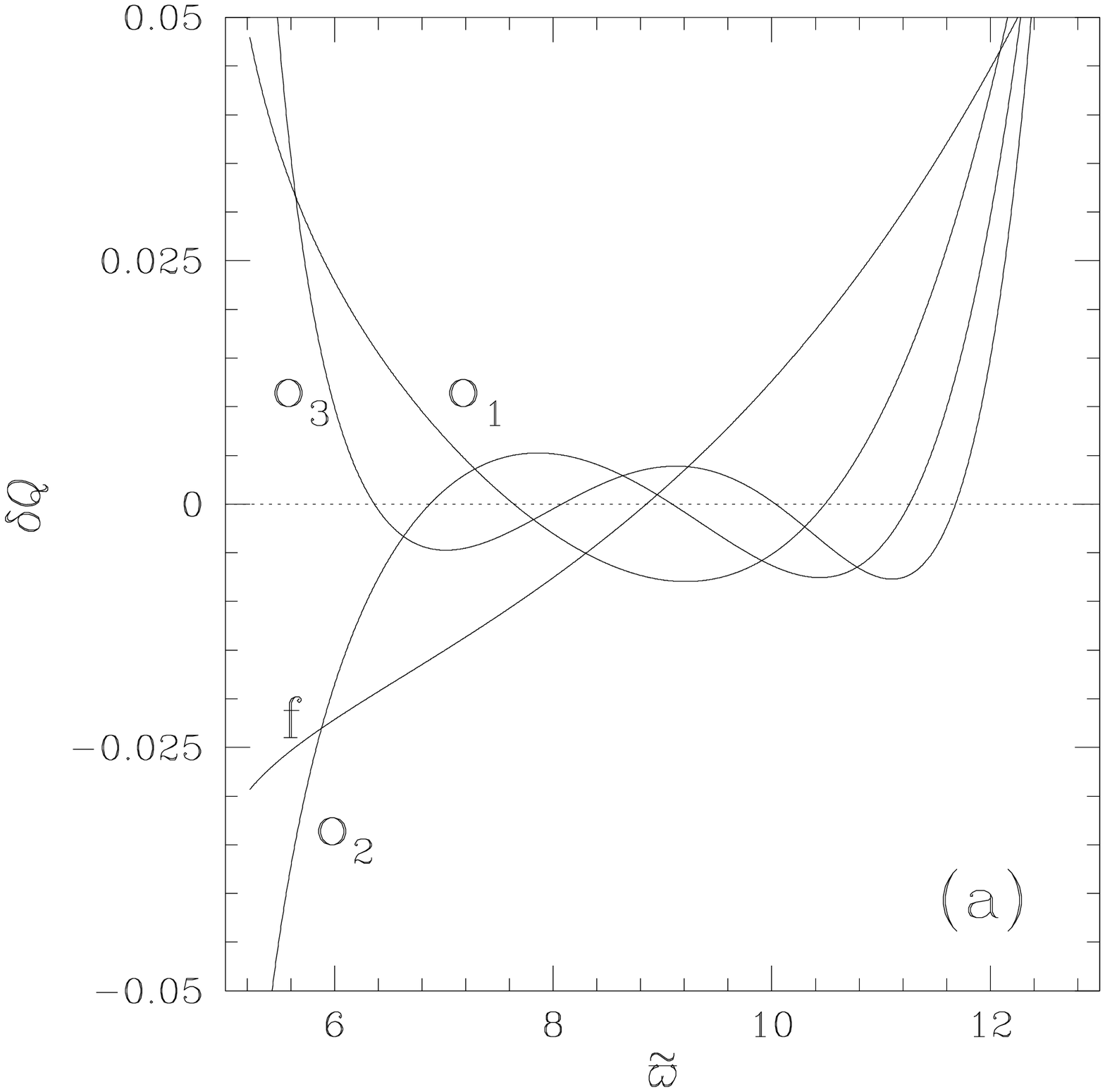,angle=0,width=8.0cm,height=7.25cm}
\hspace{0.250cm}
\psfig{file=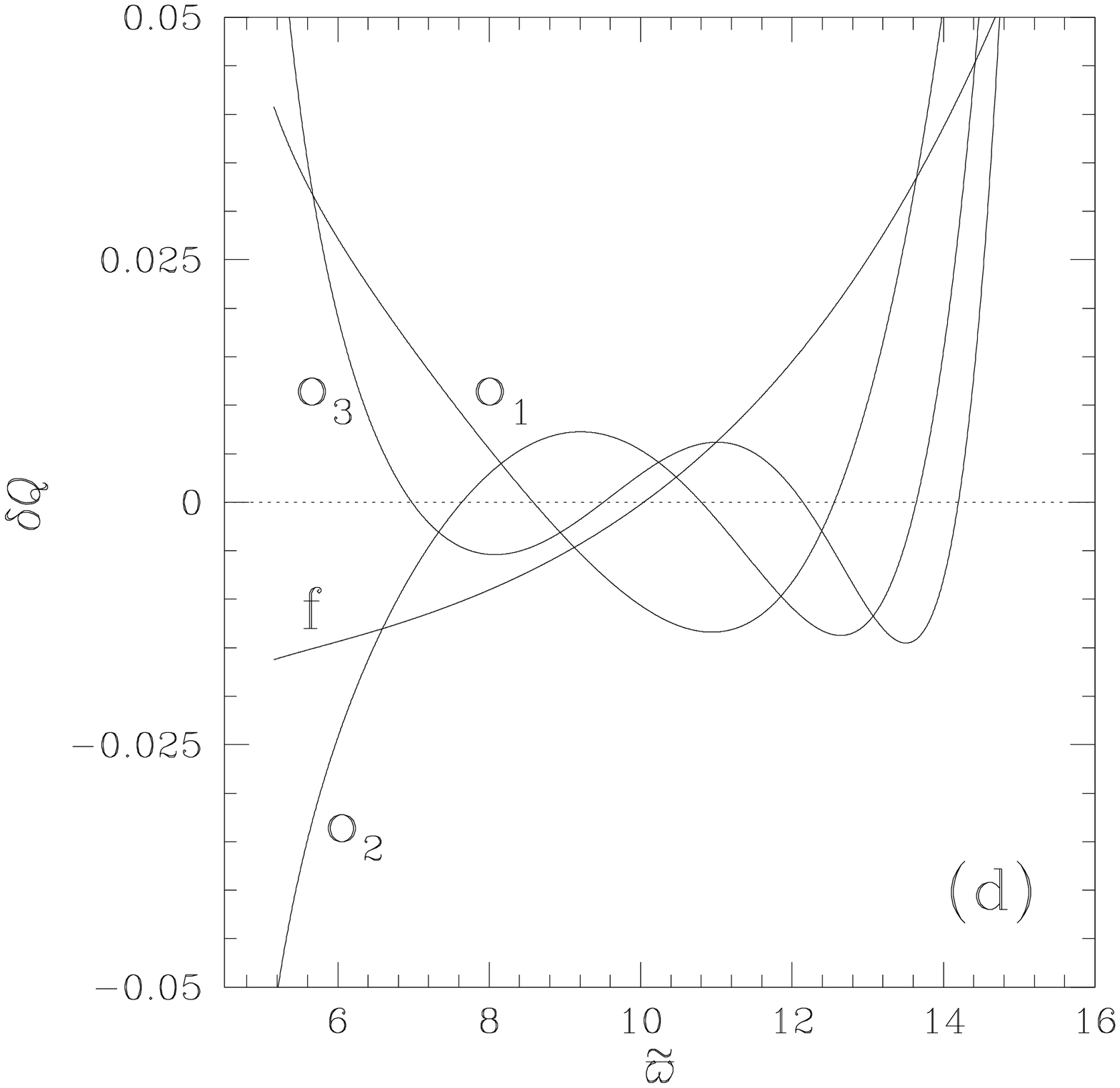,angle=0,width=8.0cm,height=7.25cm}
\vspace{0.0125cm}
\hspace{0.001cm}
\psfig{file=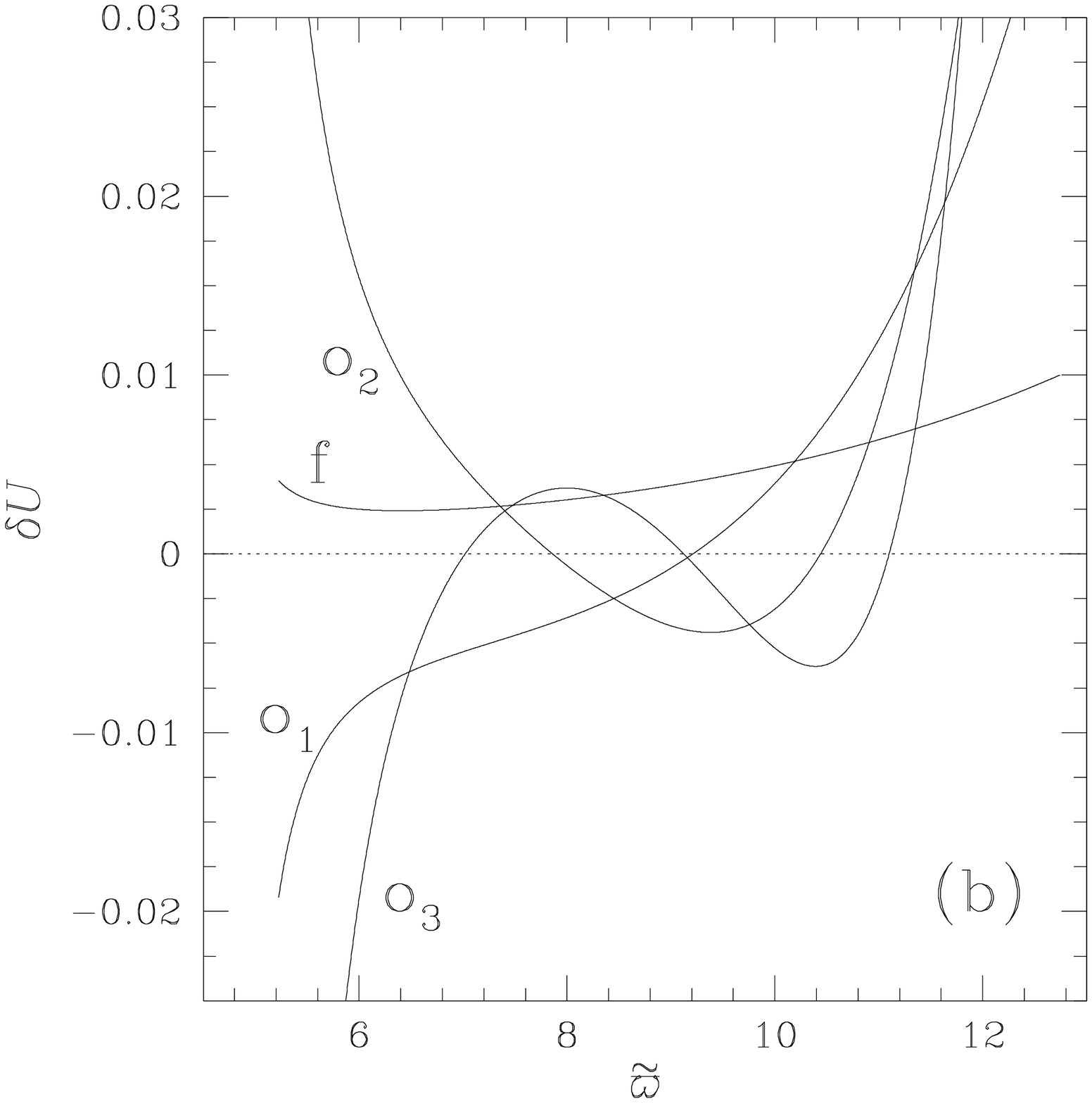,angle=0,width=8.0cm,height=7.25cm}
\hspace{0.250cm}
\psfig{file=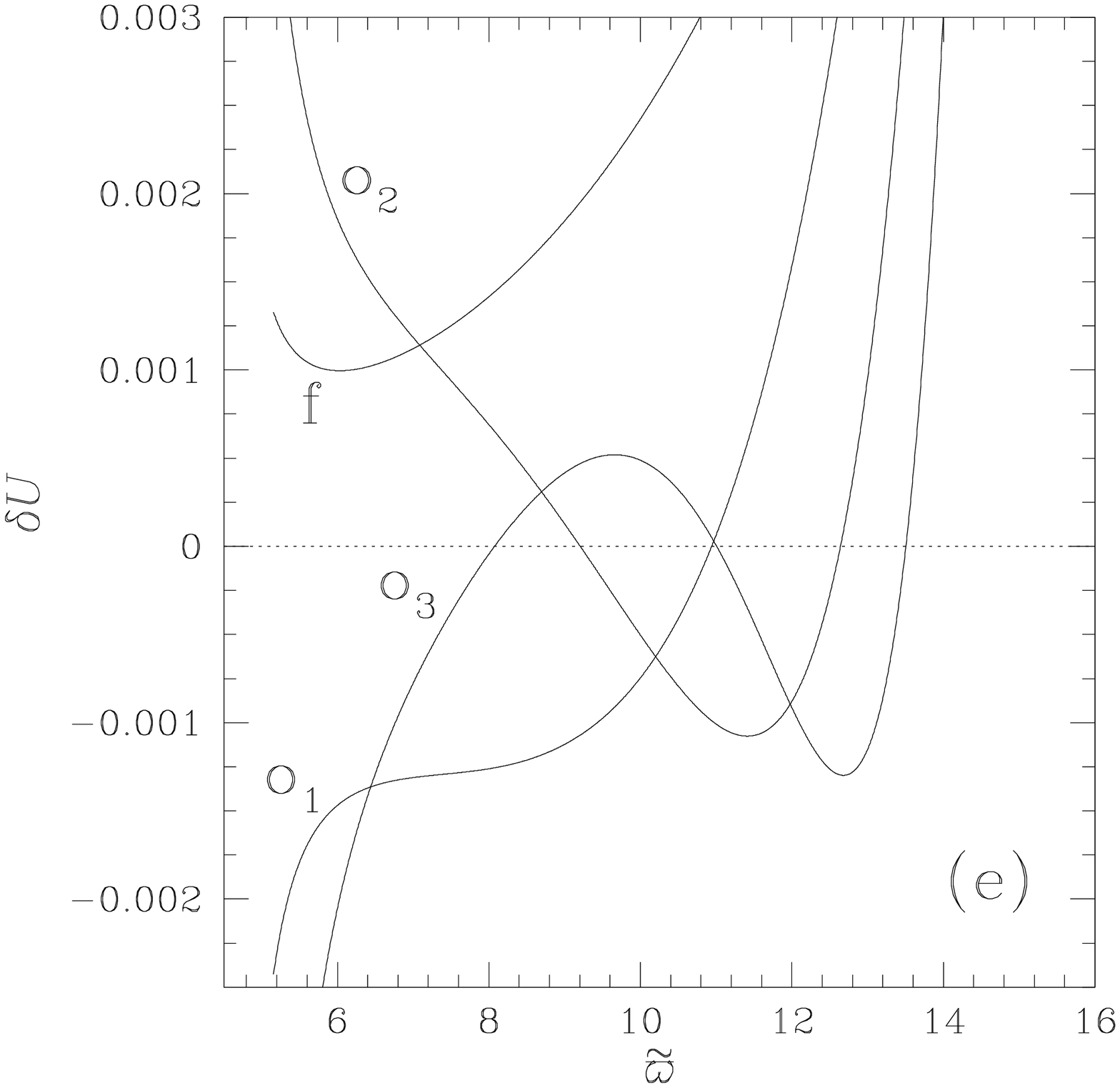,angle=0,width=8.0cm,height=7.25cm}
\vspace{0.0125cm}
\hspace{0.001cm}
\psfig{file=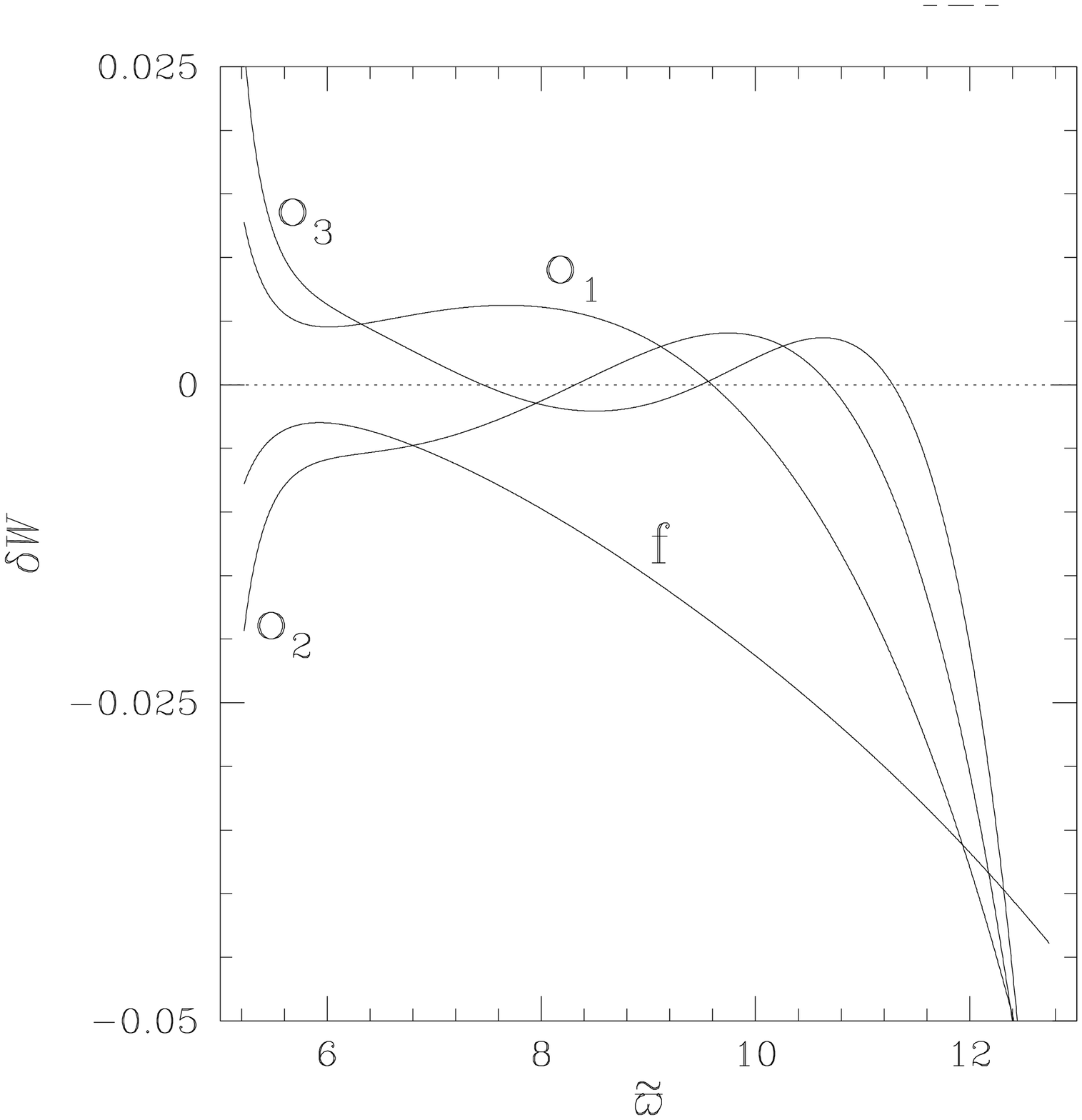,angle=0,width=8.0cm,height=7.25cm}
\hspace{0.250cm}
\psfig{file=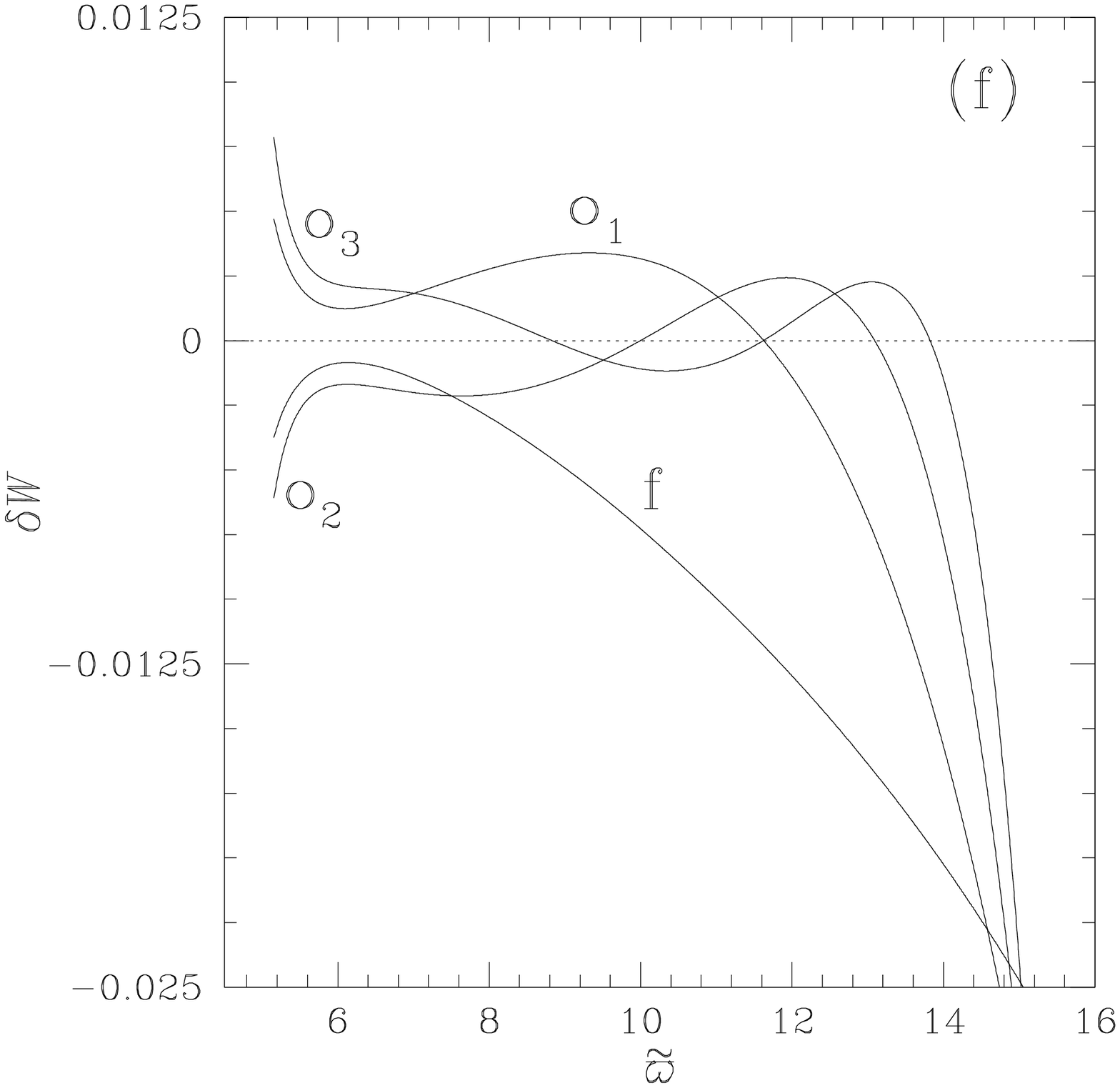,angle=0,width=8.0cm,height=7.25cm}
\end{center}
\vspace*{0.0cm}
\caption{ \small{Eigenfunctions for $\delta Q=\delta P/(E+P)$, $\delta
U=i\delta V^{\hat \varpi}$, and $\delta W=\delta V^{\hat{\phi}}$ as a
function of the radial coordinate for a tori with a ``linear'' [panels
(a)--(c); model (l1)] and a ``power-law'' [panels (d)--(f); model (pl3)]
distribution of the specific angular momentum. While the units used on
the vertical axes are arbitrary, it should be noted that the values on
panel (e) are one order of magnitude smaller than the corresponding ones
on panel (b). In all panels only the fundamental mode $f$ and the first
three overtones have been reported.}}
\label{ncsamd}
\end{figure*}

	A rapid analysis of the eigenfrequencies in Table~2 as well as of
the eigenfunctions in the various panels of Figure~\ref{ncsamd} indicates
that, overall, the results for non-constant distributions of specific
angular momentum do not differ, at least qualitatively, from those
computed for constant distributions. Most importantly, the
eigenfrequencies continue to appear in the simple sequence
\hbox{2:3:4:$\ldots$}, at least for the lowest-order modes considered
here.

	This result may appear surprising in light of the fact that the
the radial epicyclic frequency defined by (\ref{gr_epicyclic}) does not
vanish in the case of non-costant specific angular momentum
distributions, so that the computed eigenfrequencies contain also a
contribution coming from the inertial oscillations [cf. the dispersion
relation (\ref{dispersion-gr})]. However, it is important to bear in mind
that the $p$ modes discussed here behave essentially as sound waves
trapped inside the torus. As a result, while their absolute frequencies
will be different when tori with different radial extensions are
considered, the sequence in which they appear will basically follow the
relation $\sigma_n = [(2+n)/2] f$.

	The ``universality'' in which the lowest-order $p$-mode seem to
appear has an interest of its own, but may also have important
implications for the high frequency quasi-periodic oscillations observed
in binaries containing a black hole candidate (Strohmayer 2001; Remillard
et al., 2002). In at least three systems, in fact, the QPOs have been
detected with relatively strong peaks that are in a harmonic ratio of
small integers 1:2, 2:3, or 1:2:3 (Abramowicz, Kluzniak 2001), and a
possible explanation of this phenomenology in terms of the oscillations
of a geometrically thick, non-Keplerian disc of small radial extent
orbits around the black hole will be presented in a separate paper
(Rezzolla et al., 2003).

	Before concluding this Section, we should underline an important
qualitative difference that emerges when non-constant distributions of
specific angular momentum are considered. This is related with the
behaviour of the eigenfunctions for the radial velocity perturbation
$\delta U$ at the inner edge of the disc. A rapid comparison of
Figure~\ref{U_l0} with the corresponding panels (b) and (e) of
Figure~\ref{ncsamd} shows that the eigenfunctions become vanishingly
small at the inner edge of the disc in the case of ``power-law''
distribution of specific angular momentum. This does not seem to be case
for the behaviour at the outer edge, where the eigenfunctions maintain
very large values. Similarly, this behaviour at the inner edge is not
encountered with the other two distributions considered [note that the
values on vertical axis of panel (e) are one order of magnitude smaller
than the corresponding ones on panel (b) and in Figure~\ref{U_l0}].

	A physical explanation for this behaviour can be found by
comparing the behaviour of the radial epicyclic frequency curves for the
different distributions of angular momentum. The lower panel of
Figure~\ref{epic_1}, in particular, shows that the radial epicyclic
frequency for both the linear and the ``power-law'' distributions of
specific angular momentum are monotonically decreasing for increasing
radii (This has been found to hold for all of the disc models considered
here.). However, while the two frequencies tend to be comparable in the
outer regions of the torus, they differ considerably in the inner
regions, where the epicyclic frequency for a power-law distribution
increases inward much more rapidly. More importantly, the epicyclic
frequency is always {\it smaller} than the fundamental frequency for a
torus with a linear distribution of angular momentum, while this is not
the case for a power-law distribution (cf. Table~\ref{freq-lin}).
	
	In this latter case, then, there will be regions of the disc in
which $\sigma^2 - \kappa^2_{\rm r} \sim k^2 c^2_s < 0$ and thus where the
amplitude of any acoustic wave is forced to be zero (Other modes, such as
the $g$ modes, can however exist in this region, in which they are
actually trapped; Kato and Fukue, 1980). Such a region is referred to as
an {\it evanescent-wave} region and effectively represents the part of
the disc in which the the centrifugal barrier produced by rotation makes
the radial epicyclic frequency so high that all inwarding moving acoustic
waves cannot propagate but are reflected back to larger radial
positions. In those parts of the disc where the fundamental frequency is
larger than the radial epicyclic one, on the other hand, $\sigma^2 -
\kappa^2_{\rm r} > 0$, and sound waves can propagate undamped at least in
a perfect fluid. Stated differently, the results found here indicate
that, when acoustic waves are considered, the inner regions of the torus
are of evanescent type for discs with a power-law distribution of angular
momentum, while no evanescent region exists in a constant or in a linear
distribution of specific angular momentum.

	Finally, it is worth pointing out that the behaviour discussed
above can be used to interpret some of the most recent results on the
occurrence of the runaway instability (Font and Daigne 2002a, 2002b;
Zanotti et al., 2003). In a series of relativistic hydrodynamics
calculations of a torus orbiting and accreting onto a rotating black
hole, in fact, Font and Daigne (2002b) have found that while tori with
constant specific angular momentum distributions lead to a runaway
instability, a slight outward increase as a power-law of the specific
angular momentum, has a dramatic stabilizing effect, suppressing the
instability completely. While this behaviour reflects a process which is
fundamentally nonlinear, it is indeed consistent with the interpretation
that the inner regions of thick discs prevent the inward propagation of
perturbations and hence tend to suppress the accretion of mass onto the
black hole. Because the runaway instability is fed by the changes in the
spacetime produced by the mass accretion, a considerable reduction of the
latter can be the cause of the suppression of the former. A more detailed
investigation is necessary to validate this hypothesis.

\section{Conclusions}
\label{conclusions}

	We have presented the first investigations of the oscillation
properties of relativistic, non-selfgravitating tori orbiting around a
black hole. More precisely, we have here considered the axisymmetric
oscillation modes of a geometrically thick disc constructed in a
Schwarzschild spacetime. Like relativistic stars, relativistic tori are
extended objects with non-trivial equilibrium configurations governed by
the balance among gravitational, centrifugal and pressure forces. Unlike
relativistic stars, however, the contributions coming from centrifugal
forces are not small in relativistic tori and this amplifies the role of
oscillations restored by centrifugal forces. This leads to oscillations
properties that can be considerably different from those encountered in
stars.

	Because thick discs have angular momentum distributions that are
intrisically non-Keplerian, we have modeled them with a number of
different distributions of specific angular momentum, which have been
chosen to be either constant within the torus, or with a radial
dependenced being linear or a power-law. Furthermore, in order to keep
the treatment as simple as possible and handle equations analytically
whenever possible, we have built the models for the tori using vertically
integrated and vertically averaged quantities.

	Rather little is still known about the oscillation modes of
relativistic tori and our approach to the problem has therefore
progressed by steps. In particular, our first step has been that of
considering the dispersion relation for acoustic waves propagating within
these objects. We have done this firstly in a Newtowian framework in
which equations are simpler and so is their physical interpretation. We
have then extended the local analysis to a general relativistic framework
and in particular to a Schwarzschild black hole spacetime. While a local
analysis and the resulting dispersion relation is valid only for
oscillations with wavelengths small when compared with the typical
lengthscale in the tori, the ones derived here have been important to
clarify the relation between the acoustic waves and the other waves that
play a fundamental role in fluids orbiting in a central potential, i.e.
inertial (or epicyclic) waves. In particular, it has been possible to
show that, in general, both acoustic and inertial oscillations are
present in the perturbations of thick discs and that the latter can be
removed only in the special case of a constant distribution of specific
angular momentum. This result, which was already know in Newtonian
physics, has here been shown to hold also for extended relativistic fluid
configurations. To to best of our knowledge this result has never been
discussed in the literature before.

	Going a step further and beyond a local analysis, we have used
the same mathematical setup to perform a global analysis and determine
both the eigenfunctions and the eigenfrequencies of the axisymmetric
oscillations. Also in this case, the assumption of a vertically
integrated equilibrium has simplified the eigenvalue problem
considerably, translating it into a set of coupled ordinary differential
equations which have been solved using standard techniques and for a
number of different models for the tori. The modes found in this way
correspond to the $p$ modes of relativistic tori and are characterized by
eigenfrequencies appearing in a simple sequence of small integers
2:3:4:$\ldots$, at least for the first lower-order modes. This important
feature does not depend on the distribution of angular momentum used to
build the tori and basically reflects the fact that $p$ modes are to be
interpreted as sound waves trapped within the cavity represented by the
confined torus.

	The properties found here with a linear global analysis are in
very good agreement with the numerical results found in the time
evolution of perturbed ``toroidal neutron stars'' having constant
distributions of specific angular momentum (Zanotti et al.,
2003). This agreement with fully nonlinear two-dimensional simulations
provides convincing evidence that the assumption of a vertically
integrated equilibrium has not subtracted important information from the
present analysis.

	A number of motivations are behind the investigations carried in
this paper. Firstly, and as mentioned above, relativistic tori possess a
rich variety of oscillation modes which are still essentially unexplored,
in stark contrast with what is known (both in Newtonian and in
relativistic regimes) about geometrically thin discs. In this respect,
the present work was intended as a first investigation of the
discoseismology of geometrically thick discs. Secondly, our analysis was
aimed at interpreting and clarifying some of the results found by Zanotti
et al. (2003), who had investigated numerically the dynamical response of
massive tori to perturbations but had not been able to characterize all
of the mode properties. Finally, by investigating the response of tori to
perturbations, we intended to highlight possible connections between the
oscillation modes of these objects and all those astrophysical scenarios
in which geometrically thick discs may play an important role.

	Some of the results discussed here could find application in more
general contexts. One of such applications is offered by the
quasi-periodic X-ray phenomenology observed in X-ray binary systems
containing a black hole candidate (Strohmayer 2001; Remillard et al.,
2002). In these systems, in fact, the X-ray emission is modulated
quasi-periodically with power spectra having peaks in a harmonic ratio of
small integers 1:2, 2:3, or 1:2:3 (Abramowicz and Kluzniak, 2001). While
there are a number of possible explantions for this behaviour, it could
be interpreted simply in terms of the $p$-mode oscillations of a
geometrically thick, non-Keplerian disc of small radial extent orbiting
around the black hole (Rezzolla et al., 2003). Another of such
applications is offered by the runaway instability which has recently
been considered by a number of authors (Font and Daigne 2002a, 2002b;
Zanotti et al., 2003) through fully nonlinear numerical calculations. In
particular, the numerical evidence that the instability can be suppressed
if the torus has a distribution of angular momentum which is slightly
increasing outwards, can be interpreted simply in terms of the behaviour
of the eigenfunctions for the radial velocity perturbations, which become
vanishingly small in the inner regions of the disc. This, in turn, is
produced by the appearence of a region of evanescent-wave propagation in
the inner regions of the torus that prevents the inward propagation of
perturbations, reduces the accretion of mass onto the black hole, thus
suppressing the instability.

	As mentioned in the abstract, the present work represents the
first of a series of papers aimed at a systematic investigation of the
oscillation properties of relativistic tori. We are presently
investigating extensions of the present approach in which the $p$ modes
are investigated both when the background spacetime is that of a Kerr
black hole (Rezzolla \& Yoshida, in preparation), and when deviations
from axisimmetry are present. In addition to this, work is also in
progress to extend the solution of the eigenvalue problem to a fully
two-dimensional model for the torus. The results of these investigations
will be presented in future papers.

\section*{Acknowledgments}

It is a pleasure to thank N. Andersson, O. Blaes, T. Font, W. Kluzniak,
and J. Miller for many useful discussions. Financial support for this
research has been provided by the MIUR and by the EU Network Programme
(Research Training Network Contract HPRN-CT-2000-00137). The computations
were performed on the Beowulf Cluster for numerical relativity {\it
``Albert100''}, at the University of Parma.
 
\appendix

\section[]{On the values of the eigenfrequencies when 
$L\rightarrow 0$}

In the limit of vanishing torus sizes, $L\rightarrow 0$, all of the
eigenfunctions can be assumed to have only a simple linear dependence on
$\varpi$, i.e.
\begin{eqnarray}
&&\delta U = {\rm a}\varpi+{\rm b} \ , 	\\
\nonumber 			\\
&&\delta W = {\rm c}\varpi+{\rm d} \ , 	\\
\nonumber 			\\
&&\delta Q = {\rm e}\varpi+{\rm f} \ , 
\end{eqnarray}
where ${\rm a, b},\ldots, {\rm f}$ are just constant coefficients. Using
this ansatz in the system of perturbed equations
(\ref{euler-r})-(\ref{cont_gr}), we obtain the following system of
equations
\begin{eqnarray}
&& \varpi[\sigma {\rm a}+2{\rm c} e^{-\lambda}\Omega S(\varpi) + i k A
	e^{-\nu-\lambda} {\rm e}] + \label{cond_1}
\nonumber \\
&& \hskip 1.5truecm \sigma {\rm b} + 2 {\rm d} e^{-\lambda} \Omega
	S(\varpi) + ik A e^{-\nu-\lambda}{\rm f}=0 \ , \\
\nonumber \\
&& \varpi[\sigma {\rm c} + {\rm a} e^{-\lambda} H(\varpi)] + 
	\sigma {\rm d} + {\rm b} e^{-\lambda} H(\varpi) =0 \ , \\
\nonumber \\
&& \varpi\left[\sigma {\rm e} + \frac{\Gamma P}{E+P} {\rm a} 
	e^{\nu-\lambda} i k\right] + \sigma {\rm f} 
\nonumber \\
&& \hskip 3.5truecm +\frac{\Gamma P}{E+P}e^{\nu-\lambda} i k {\rm b}
	=0 \label{cond_3} \ ,
\end{eqnarray}
where 
\begin{eqnarray}
&& H(\varpi)\equiv 2\Omega + \varpi\frac{d\Omega}{d\varpi} - 
	2 \Omega \varpi \frac{d\nu}{d\varpi} \ , 
\\ \nonumber \\
&& S(\varpi)\equiv 1+\frac{\varpi}{E+P}\frac{dP}{d\varpi}\ , 
\end{eqnarray}
and where it is easy to realize that the radial epicyclic frequency is
then simply given by [cf. eq. (\ref{gr_epicyclic})]
\begin{equation}
\kappa^2_{\rm r} = 2 e^{-2\lambda}\Omega H(\varpi) S(\varpi) \ .
\end{equation}

	Each of the equations in the system
(\ref{cond_1})--(\ref{cond_3}) can be written symbolically as $\alpha_i
\varpi + \beta_i=0$, $i=1,2,3$ with $\alpha_i$ and $\beta_i$ being just
shorthands for the longer expressions in
(\ref{cond_1})--(\ref{cond_3}). Furthermore, because each equation in of
the (\ref{cond_1})--(\ref{cond_3}) must hold for any $\varpi$, we need to
impose that all of the coefficients $\alpha_i$ and $\beta_i$ vanish
simultaneously. This generates a system of six equations in the six
unknowns ${\rm a, b,\ldots, f}$
\begin{eqnarray}
\sigma {\rm a} + 2{\rm c}\Omega e^{-\lambda} S(\varpi) + ik A
	e^{-\nu-\lambda} {\rm e} &=&0 \ , \label{sys_1}
\\\nonumber \\
\sigma {\rm b} + 2 {\rm d} \Omega e^{-\lambda} S(\varpi) + ik A
	e^{-\nu-\lambda} {\rm f} &=&0 \ , 
\\\nonumber \\
\sigma {\rm c} + {\rm a} e^{-\lambda} H(\varpi) &=&0 \ , 
\\\nonumber \\
\sigma {\rm d} + {\rm b} e^{-\lambda} H(\varpi) &=&0 \ , 
\\\nonumber \\
\sigma {\rm e} +\frac{\Gamma P}{E+P} e^{\nu-\lambda} i k {\rm a}
	&=&0 \ , \label{sys_6} 
\\\nonumber \\
\sigma {\rm f} + \frac{\Gamma P}{E+P} e^{\nu-\lambda} i k {\rm b} &=&0 \ ,
\end{eqnarray}
which, in matrix form, can be written as
\begin{equation}
{\cal A} 
\left(\begin{array}{c}
	{\rm a} 	\\ \nonumber \\ 
	{\rm b} 	\\ \nonumber \\ 
	\vdots 		\\ \nonumber \\ 
	{\rm f} 
\end{array}\right)
	= 0 \ .
\eqno	(A16)
\end{equation}

	A non-trivial solution to the system (A16) is possible if the
determinant of the matrix ${\cal A}$ vanishes and, after lenghty but
straightforward calculations, this condition yields
\begin{eqnarray}
\setcounter{equation}{17}
\label{detaeq0}
&& {\rm det}({\cal A}) =\sigma^2\left[ \sigma^2-2\Omega 
	e^{-2\lambda} H(\varpi)	S(\varpi)\right]
	\biggl[ \sigma^2 -
	\nonumber \\
&& \hskip 0.5 truecm
	2 \Omega e^{-2\lambda} H(\varpi)S(\varpi) +
	\sigma\left(\frac{\Gamma P}{E+P}\right)^2 k^4 A^2 e^{-4\lambda}+ 
	\nonumber \\
&& \hskip 3.5 truecm
	2\frac{\Gamma P}{E+P}k^2 A e^{-2\lambda}
	  \biggr] = 0 
\end{eqnarray}
Since ${\Gamma P}/{E+P}\sim c_s^2$, all of the terms proportional to this
quantity can be neglected in (\ref{detaeq0}), which then reduces to
\begin{equation}
\label{sigma_kappa}
\sigma^2\left[\sigma^2-2\Omega 
	e^{-2\lambda} H(\varpi)S(\varpi)\right]^2=0 \ .
\end{equation}
A non-trivial solution of equation (\ref{sigma_kappa}) is then given 
by 
\begin{equation}
\sigma^2  = 2\Omega e^{-2\lambda}
	H(\varpi)S(\varpi) = \kappa_{\rm r}^2 \ ,
\end{equation}
thus proving that the eigenfrequencies tend to the value of the epicyclic
frequency in the limit ot $L\rightarrow 0$.

\section[]{On the sound speed in rotating polytropic tori}

	We here show that a {\it non-self gravitating, polytropic} fluid
configuration orbiting in {\it hydrostatic equilibrium} around a
Schwarzschild black hole has a sound velocity which is invariant under
changes of the polytropic constant $k$. To prove this consider the
equation of hydrostatic equilibrium (\ref{bernoulli}) which, in the
simplified metric (\ref{eqtrl_metric}), can be rewritten as
\begin{equation}
\label{hydrostat}
\frac{\partial_\varpi p}{e+p}=-
	\frac{e^{2\nu} \partial_{\varpi}\nu - 
	\Omega^2\varpi}{e^{2\nu} - \Omega^2\varpi^2} 
	= f(\varpi)\ .
\end{equation}
The right-hand-side of equation (\ref{hydrostat}) depends only on the
kinematic and geometric properties of the disc, namely on the angular
momentum distribution (through $\Omega$) and on the external
gravitational field of the central black hole (through $\nu$). Both of
these quantities depend on $\varpi$ only.

	Introducing the relation $\gamma=1+1/n$, the polytropic EOS
\hbox{$p = k\rho^\gamma$} can also be written in terms of the Emden
function \hbox{$\Theta^n \equiv \rho$} as
\begin{equation}
p = k \Theta^{N+1}\ , 
\end{equation}
so that equation (\ref{hydrostat}) effectively becomes
\begin{equation}
\label{psi_eq}
\frac{\partial_{\varpi}\psi}{1+\psi} = f(\varpi)\ ,
\end{equation}
where we have defined $\psi \equiv k(n+1)\Theta$.

	Equation (\ref{psi_eq}) can be integrated analytically to give 
\begin{equation}
\label{integral}
\psi(\varpi) = {\cal C} \exp\left[\int_{\varpi_{\rm in}}^\varpi 
	f(\varpi')d\varpi'\right] - 1\ ,
\end{equation}
where ${\cal C}$ is an arbitrary constant. 

	The important result contained in equation (\ref{integral}) is
that \hbox{$\psi$} is a function of $\varpi$ only and will not depend,
therefore, on the specific value chosen for $k$. As a result, once
$\gamma$ (and thus $n$) is fixed, any transformation $K\rightarrow
K\alpha$, where $\alpha$ is an arbitrary constant, must be accompanied by
a corresponding transformation $\Theta \rightarrow \Theta/ \alpha$. An
important consequence of this is that all of the quantities given by
$p/\rho$, $dp/d\rho$, $p/e$, $dp/de$ are invariant under changes in
$k$. Equally invariant is the sound speed defined as $c_s \equiv
\sqrt{dp/de}$.

	To appreciate the physical implications of this result, it is
useful to recall that the polytropic constant plays here the role of
determining the total rest-mass of the torus. Once the other parameters
of the background model have been fixed (i.e. the distribution of angular
momentum and the potential gap), in fact, the rest-mass of the torus is
calculated from the integral
\begin{equation}
\label{rest_mass}
M_{\rm t} \equiv \int_{\varpi_{\rm in}}^{\varpi_{\rm out}} 
	2\pi \varpi {\Sigma} u^t \varpi d\varpi \ . 
\end{equation}
As a consequence, different choices for $k$ will yield tori with
different density distributions $\Sigma$ and hence rest-masses, while
maintaining the same radial extension. The result contained in equation
(\ref{integral}) states, therefore, that it is possible to build tori
with largely different rest-mass densities and yet have them all share
the same sound velocity.
	
	Two additional remarks are worth making. The first one is that
while this result has been derived for a disc with a nonzero vertical
extension, it applies also for a vertically integrated model. The second
remark is that while we have been here concentrating on relativistic
models, the same result can be shown to apply in Newtonian physics when
the central black hole is replaced by a generic spherically symmetric
gravitational potential.


\label{lastpage}  
\end{document}